\documentclass[11pt]{book}

\usepackage{epsfig}
\usepackage{fancyhdr}
\usepackage{amsfonts}
\usepackage{amssymb}
\usepackage{amsbsy}
\usepackage{latexsym}

\newcommand{\no}{{\noindent}}

\newcommand{\C}{{\mathbb C}}

\newcommand{\myphantom}[1]{\rule{0pt}{#1}}

\newcommand{\comment}[1]{}

\def\bbbz{{\sf Z\!\!\!Z}}
\def\sl2z{SL(2,\bbbz)}
\newcommand{\be}{\begin{equation}}
\newcommand{\ee}{\end{equation}}
\newcommand{\bea}{\begin{eqnarray}}
\newcommand{\eea}{\end{eqnarray}}

\def\bbbz{{\sf Z\!\!\!Z}}
\def\sl2z{SL(2,\bbbz)}

\def\z0{{\bf z_0}}

\newcommand{\bit}{\begin{itemize}}
\newcommand{\eit}{\end{itemize}}
\newcommand{\gtt}{\sqrt{- G_{tt}}}
\begin{document}

%Title Page

\newpage
\pagestyle{empty}
\begin{flushright}
USITP-03-04\\
hep-th/0304143
\end{flushright}

\vspace*{8mm}

\begin{center}
  {\Huge {\bf If I Only Had A Brane!} }\\ 
\vskip.5cm
 {\huge {\cal Tasneem Zehra Husain}} 
\end{center}
\vskip1cm

\begin{figure}[ht]
\epsfxsize=9cm
\centerline{\epsfbox{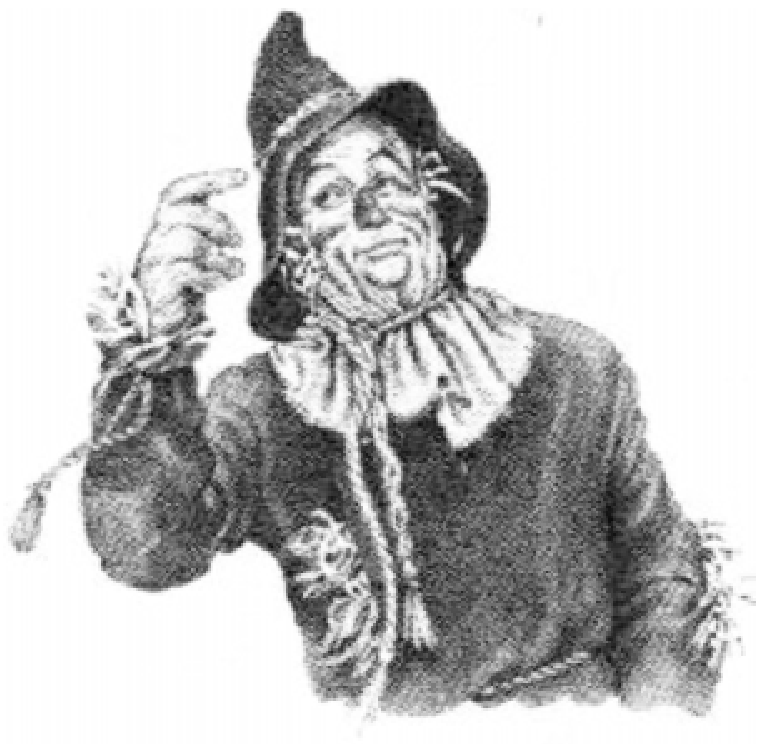}}
\end{figure}

\vskip1cm

\begin{figure}[ht]
\epsfxsize=2cm
\hskip4cm
{\epsfbox{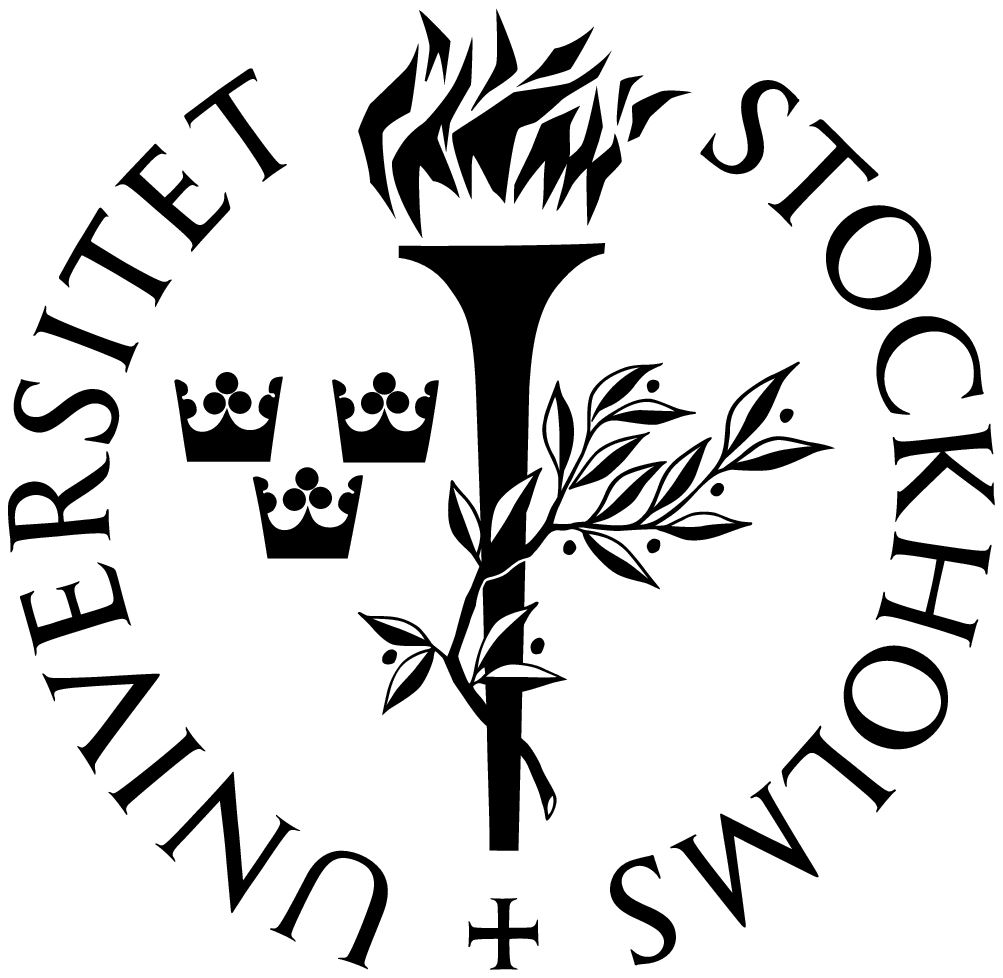}} 
{\epsfxsize=4cm \hskip.5cm {\epsfbox{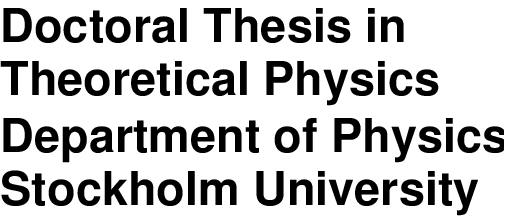}}}
\end{figure}

%ISBN

\clearpage

\vspace*{\fill}

\noindent
Thesis for the degree of Doctor of Philosophy in Theoretical Physics\\
Department of Physics, Stockholm University\\
Sweden\\[\baselineskip]
\copyright \ Tasneem Zehra Husain, 2003.\\
ISBN 91-7265-592-5 (pp. 1 - 65)\\
Akademitryck AB, Edsbruk.

%Dedication 

\newpage

\myphantom{30mm}
\begin{center}
{\it \large To Nanna, who would be very proud \\ 
and Baji who would be really glad it's finally over!} 
\end{center}

\newpage
\clearpage
\myphantom{45mm}

\newpage

\subsection*{The Wizard Of Oz Theme}

For some inexplicable reason, this thesis has taken on
a life and a theme of its own. It started with the title I guess and
then everything else just followed suit. As a result, there are
references to 'The Wizard Of Oz' all over the place. For those of you 
who have not seen the classic 1931 MGM musical\footnote{Though the
movie is based on a book by L.Frank Baum, the images and quotes used 
here are from the film.} starring Judy Garland, here is a quick synopsis.

Dorothy is a young girl who lives with her Aunt and Uncle on a farm 
in Kansas. She is on her way home when a huge tornado sweeps the
land but by the time she reaches the farm-house, everyone else is already
safe in the storm shelter underground. The tornado blows the house 
away carrying
Dorothy and her dog Toto inside it. When the house finally lands, 
Dorothy opens the door and steps out into a magical world where things
are unfamiliar, but beautiful. The film which started out in sepia
bursts into Technicolour at this point as Dorothy utters one of the
most famous sentences of the film, {\em ``Toto, I have a feeling we're not 
in Kansas anymore. We must be over the rainbow!''}. The song {\bf Over the
Rainbow} plays in the background.

Before Dorothy has had time to look around 
in this strange new place, a shimmering ball floats in, carrying the
lovely Glinda, the Good Witch of the North. Glinda announces to
Dorothy that her house has landed on, and killed, the Wicked Witch 
of the East who had ruled over Munchkinland -- the wonderful country
where Dorothy finds herself now. All that remains of the Wicked Witch
are her feet which stick out from
under the house with their sequin encrusted Ruby slippers. The
jolly dwarf-like Munchkins, having quite suddenly and unexpectedly 
been liberated from a tyrannical reign 
burst into joyous celebration. Their cheer is
brought to an abrupt end with the arrival of the Wicked Witch of the 
West who comes to swear vengence on Dorothy for killing her sister --
and to collect the Ruby slippers. When she turns to them
however, they are magically transferred to Dorothy's feet. 
The witch tries to convince Dorothy to hand the slippers over, 
saying that they will be no use to her .. but Glinda tells her to hang
on to them. They must be powerful, she says, if the Wicked Witch wants
them so badly. Thwarted at her efforts, the Wicked Witch renews her
menacing warning and disappears.

Fearful now, Dorothy expresses her desire to 
return home for safety, {\em "Which is the way back to Kansas? I can't
 go the way I came"}. Glinda suggests that she travel to the far-off 
Emerald City in the Land of Oz, since 
{\em "The only person who might know would be the great and wonderful 
Wizard of Oz himself."} To get there, Dorothy is told 
{\em "It's always best that you start at 
the beginning, and all you have to do is follow the Yellow Brick Road,"}
The Munchkins guide Dorothy to the border of 
Munchkinland to start her on her journey as they bid her farewell
singing {\bf Follow the Yellow Brick Road} and Glinda reminds her, 
{\em Never let those ruby
slippers off your feet for a moment, or you will be at the mercy of
the Wicked Witch of the West.}

Dorothy and Toto then begin the long walk to the 
Emerald City, along
the Yellow Brick Road. Quite soon they come across a fork in the road
and when Dorothy wonders out loud where to go, she gets her answer
from a talking Scarecrow! The Scarecrow, upon finding out Dorothy's
mission asks if he can join her on her quest to ask the Wizard for a
brain. He has no brain he says, which is why the crows are not scared
of him at all. In order to explain his predicament, he sings the song
{\bf If I only had a brain}. Dorothy is only too glad to have him go
along. On their journey down the Yellow Brick Road, Dorothy and the
Scarecrow are joined by a Tin Woodman, who is in need of a heart and a 
Cowardly Lion who wants the Wizard to give him some courage.

After a few adventures and set-backs, the friends reach the glittering
towers of {\bf the Emerald City}. From the outside the Emerald City looks
like a huge shimmering castle and the four friends feel sure they will
obtain their hearts' desires in such a wonderous place. Getting in to
the City does not prove an easy task however, as there is only one
door and that is bolted shut. After many attempts at knocking, the
four friends finally manage to attract the attention of the
gate-keeper who talks to them through a small window which he
reluctantly opens. The following exchange then takes place.\\

\no
Dorothy: We want to see the Wizard.

\no
Gateman: The Wizard? But nobody can see the great Oz. Nobody's
ever seen the great Oz. Even I've never seen him.

\no
Dorothy: Well then, how do you know there is one?

\no
Gateman: Because he, uh..., you're wasting my time.\\

Finally Dorothy hits upon the happy notion of saying that it was
Glinda who sent them there. As if that name was the magic password he
had been waiting for, the gate-keeper flings open the doors and lets
Dorothy and her friends in to the charming and very unique Emerald
City. After being feted and introduced to the wonders of the Emerald
City, Dorothy tries to seek an audience with the Wizard. Once again
she meets with incredulous faces. The Guard outside the Wizard's room
puts it best when he says {\em ``Orders are, nobody can see the Great
Oz, not nobody, not no how...not nobody, not no how!''}. Once again,
Dorothy has to explain who she is, and that the Wicked Witch of the
West is after her, in order to be let in.

The meeting with the Wizard is brief to say the least. The little band
of travellers is told that it is ofcourse in the Wizard's power to
grant their wishes, but first they must prove to him that they are
worthy by performing a 'small task'. They must bring to him, the
broom-stick of the Wicked Witch of the West. High drama follows 
including a scary run in with the evil Winged Monkeys, before the
friends succeed in capturing the sought after broomstick -- by pouring
cold water on the Witch and melting her to death!

With the broomstick in hand, the group returns victorious to the
Emerald City, confident that their desires will at least be
fulfilled. This time, they are allowed to meet the Wizard with no
further ado. However, when the Wizard is at his most grand and
mysterious, a little accident happens. Toto the dog, sees a curtain
and runs towards it, pulling it playfully --- and revealing a little
man who sits there, like a pupeteer controlling the illusion that is 
'The Wizard Of Oz'.  The true story now unfolds. It turns out that the
man is a magician (also from Kansas) who landed in Oz one day when his
hot air balloon flew astray. Since he came 'from the sky' he was
heralded in Oz as a Wizard. Rather than fight this myth, he furthered 
it and established himself in the Emerald City keeping the legend
alive by distancing himself from all people and not allowing his
authority to be questioned. 

He is however, a good hearted man, who only became reconciled to a
life in Oz when he thought there was no way for him to return home. He
points out to the friends how they already have what they thought they
were lacking -- all that they need now is material proof. The Lion has
displayed great courage, and in recognition is presented a medal. The
Tinman has shown great feeling and is gifted a real 'beating' heart
as proof. The scarecrow has shown great intelligence and is awarded a 
ThD (Doctor of Thinkology) degree as a result. The only person who
the Wizard can not help, is Dorothy. 

At this point, just when Dorothy starts feeling rather deflated,
Glinda shows up again. She proves that Dorothy too, already has what
she needs to make her dreams come true: the {\bf Ruby Slippers}. The 
magic, it seems, is that the person who wears the slippers can 
go where-ever they want to, as long as they {\em Click the 
heels three times and tell the slippers where to go}. 
Dorothy does this, repeating three times, {\em 
there's no place like home..} and she finds herself back at the farm. 

\newpage
\clearpage
\myphantom{45mm}

\newpage
\pagestyle{empty}

\myphantom{45mm}
\begin{center}
{\large \it
Oh I could tell you why\\
The ocean's near the shore\\
I could think of things\\
I'd never thunk before\\
And then I'd sit ...\\
And think some more.\\}
\end{center}
\vskip2cm
\hskip7cm{\sf - If I only had a Brain}

\hskip7.4cm{\bf \sf The Wizard of Oz.}

%Abstract

\newpage

\myphantom{30mm} 
\subsection*{Abstract}

\parbox{110mm}{This thesis starts with a review of supersymmetric
solutions of 11-dimensional supergravity; in particular flat M-branes and 
BPS configurations which can be constructed from them. The harmonic
function rule is discussed and it is shown why this cannot be expected
to apply to intersecting brane configurations where the intersection
is localised. A new class of spacetimes is needed to cover these
situations and the Fayyazuddin-Smith metric ansatz is introduced as 
the answer to this problem. The ansatz is then used to find
supergravity solutions for M-branes wrapped on holomorphic
curves. This method is discussed in detail and its various components
explained; for instance, the way holomorphicity dictates the
projection conditions on the Killing spinor and the construction 
of a spinor in terms of Fock space states. The method is 
illustrated via various examples. We then move on and discuss a
mathematical concept known as calibrations. Calibrations are forms 
which can be used to pick out the minimal surfaces in a given
background. Since BPS configurations of M-branes are minimal, it
turns out that they must wrap cycles 'calibrated' by such
p-forms. For the case when the background contains no flux,
calibrations have been classified. However, for more general cases 
where there is a non-trivial field strength, such a classification 
does not yet exist. This would be desirable for various reasons, one
of which is the following. Given a particular calibrated form, there is a
simple and very elegant method of writing down the supergravity
solution for a brane wrapped on the corresponding calibrated cycle. So
far, this method was applied only to Kahler calibrations as these were 
the only ones known to exist in backgrounds with non-trivial
flux. We extend this method to a wider class
which contains Kahler calibrations. A constraint is used to classify
possible calibrations; this constraint incorporates the geometry of 
the space transverse to the submanifold which contains the
supersymmetric cycle. A rule is given which can generate the required
constraint for any given M-brane wrapped on a holomorphic cycle. 
Ways in which this constraint can be satisfied are also discussed.}

%Acknowledgements

\newpage
\pagestyle{empty}

\myphantom{15mm}
\subsection*{Acknowledgments}

\parbox{110mm}{My ``Esteemed Advisor'', Ansar Fayyazuddin, took
considerable time off from contemplating 'the logic of late capitalism' 
to tend to his advisory duties. Regardless of the nature of the problem 
I pose him with, Ansar always manages to rise to the occasion.
He has taught me much, but more than that I owe him gratitude 
for giving me back my confidence at a time when Grad School was beating 
it down. I feel privileged to know him both as a student and as a friend.\\ 

It has been great fun having Fawad Hassan around. This thesis owes a 
lot to his endless supply of patience, chocolates and encouragement. 
He is always on hand 
when questions need to be answered and, together with Ioanna Pappa,
is an excellent partner in crime when some crazy scheme is being
hatched. \\  

The String Theory group here is rather like a library of friendly human 
encyclopedias. All queries can be addressed by walking down the corridor
and knocking on the doors of Ulf Lindstrom, Bo Sundborg and earlier, Alberto 
Guijosa, and Subir Mukhopadhyay. \\

Our all-girl group of students has to be a first -- on this planet at least! 
I count on Cecilia Albertsson and Teresia Mansson for various things, 
from bailing me out of trouble with the printer, to a friendly chat in 
the middle of the day; when panic hits and reassurance is needed, 
Cecilia can be found at the desk next to mine. But before it was just 
us girls, Bjorn Brinne and Maxim Zabzine were around as well. I learnt 
a lot from
both of them and still count on Bjorn for assorted help and advice.\\ 

Were it not for Britta Schmidt, I would be buried 
somewhere under piles of paper-work, with my Swedish dictionary in hand. 
She is truly the Guardian Angel of the Physics Department and I don't know 
what I would have done without her smiling face to turn to.\\ 

Thanks are also due to the Theory Group at UT-Austin for their hospitality
during my visit there. In particular, Jan Duffy for her
welcoming warmth, Julie Blum for many useful discussions and Amer Iqbal for
several fascinating conversations about Physics and everything else. }

\myphantom{15mm}

\parbox{110mm}{

I couldn't go without expressing my heartfelt gratitude to Ishrat Phophi, 
who never let me feel the lack of family in Sweden and to Khalu Jan 
who got me started on my journey here.\\ 

It's amazing what friends can pull you through. Over the course of this 
PhD, many potential therapy sessions were averted by phone calls, emails 
and visits to Annie, Aisha, Afi, Sabir, Sidra, Erej, Hajira, Mahrukh,
Aneeka, Roshani, Stella, Nicole, Saleha, Sherezade, Cristine and the 
HEPlons.\\ 

Never has the world known kids more precious than Manni and Ali. They are
the best cheer-upers on Earth with their undying faith in me and their 
unconditional love. Amir and Zaib have been adorable specially during my 
time in Austin. Despite their firm conviction that I really do nothing and 
this whole PhD thing is just an elaborate hoax they have been rather 
indulgent of it all.\\ 

Shabeeb has suffered more than his fair share at the hands of my 
PhD -- he has borne it well, with unwavering love and support. Maybe, 
now that the thesis is done, his suffering will be over as well .... 
no promises though! ;)\\

Most of all, my love and my deepest gratitude is for Abbu and Ammi, who gave 
me the roots which steady me and wings with which to fly. Between them, 
they have created such a haven that no matter where I go ``{\em
there's no place like home''}.
}
\begin{figure}[ht]
\epsfxsize=4cm
\hskip8cm
{\epsfbox{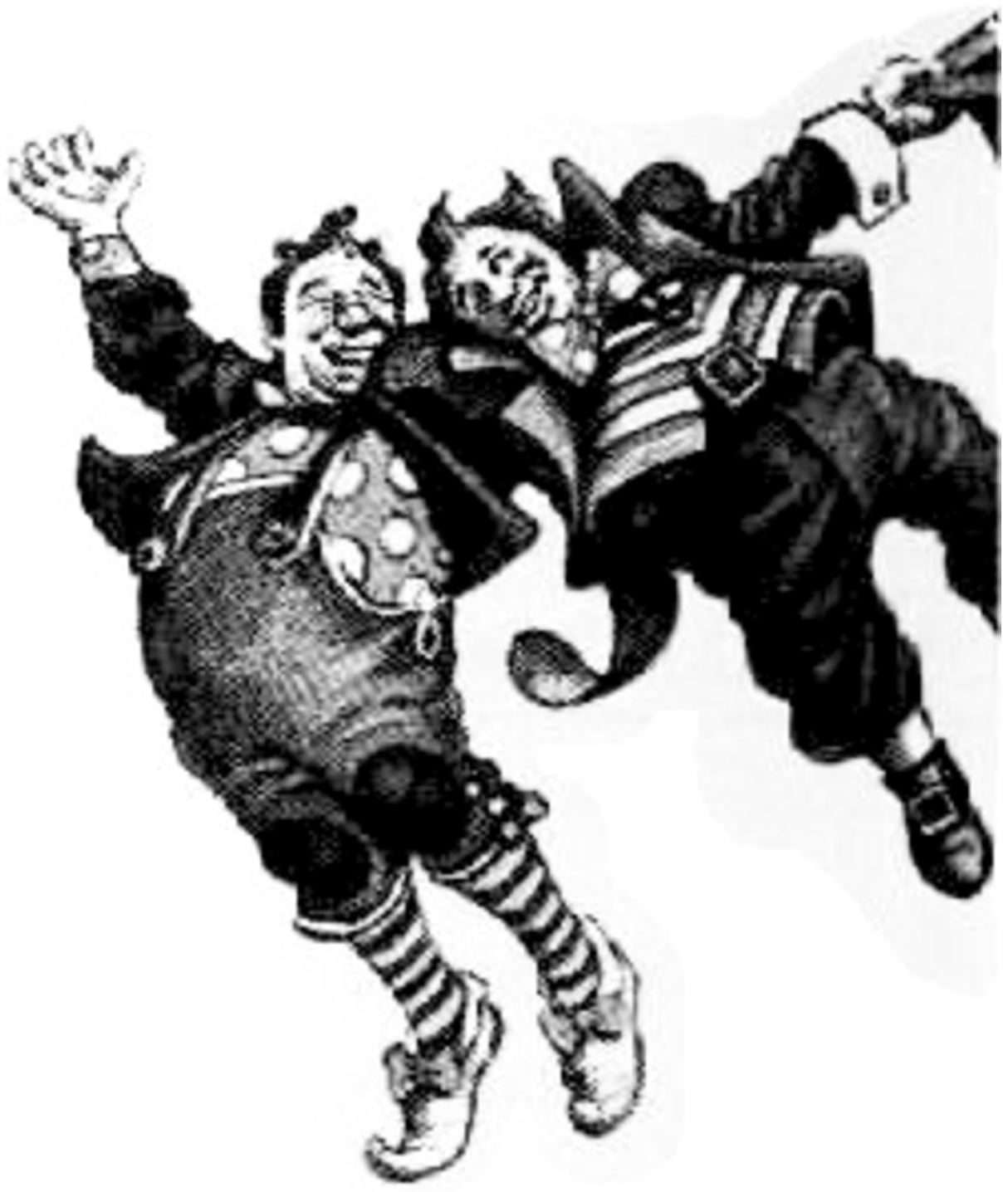}}
\end{figure}

% Contents

\addtocontents{toc}{\protect\thispagestyle{empty}} 
\tableofcontents

%\clearpage
%
%\section*{\hspace{-1.2cm}The Papers}
%
%\protect\thispagestyle{empty}
%
%\hspace{-1.2cm}
%\vbox{
%\begin{enumerate}
%\item
%${\cal N} = 1$ M5-brane geometries\\
%Bj\"orn Brinne, Ansar Fayyazuddin, Tasneem Zehra Husain, Douglas
%J. Smith,\\ 
%hep-th/0012194, {\bf JHEP 03(2001)052}
%\item
%M2-branes wrapping Holomorphic Cycles\\ 
%Tasneem Zehra Husain,\\
%{\bf hep-th/0211030}
%
%\item
%That's a Wrap!\\
%Tasneem Zehra Husain,\\
%{\bf hep-th/0302071}
%\end{enumerate}}

\newpage
\clearpage
\myphantom{45mm}

\newpage
\pagestyle{fancy}
\renewcommand{\chaptermark}[1]{\markboth{#1}{}}
\renewcommand{\sectionmark}[1]{\markright{\thesection\ #1}}
\fancyhf{}
\fancyhead[LO,RO]{\bfseries\thepage}
\fancyhead[LO]{\slshape\rightmark}
\fancyhead[RE]{\bfseries\leftmark}
\renewcommand{\headrulewidth}{1.2pt}
\renewcommand{\footrulewidth}{0pt}
\fancypagestyle{plain}{
\fancyhead{}
\renewcommand{\headrulewidth}{0pt}
}

\pagenumbering{arabic}

\chapter{Over The Rainbow}
\setcounter{page}{1}

\begin{figure}[ht]
\epsfxsize=6cm
\centerline{\epsfbox{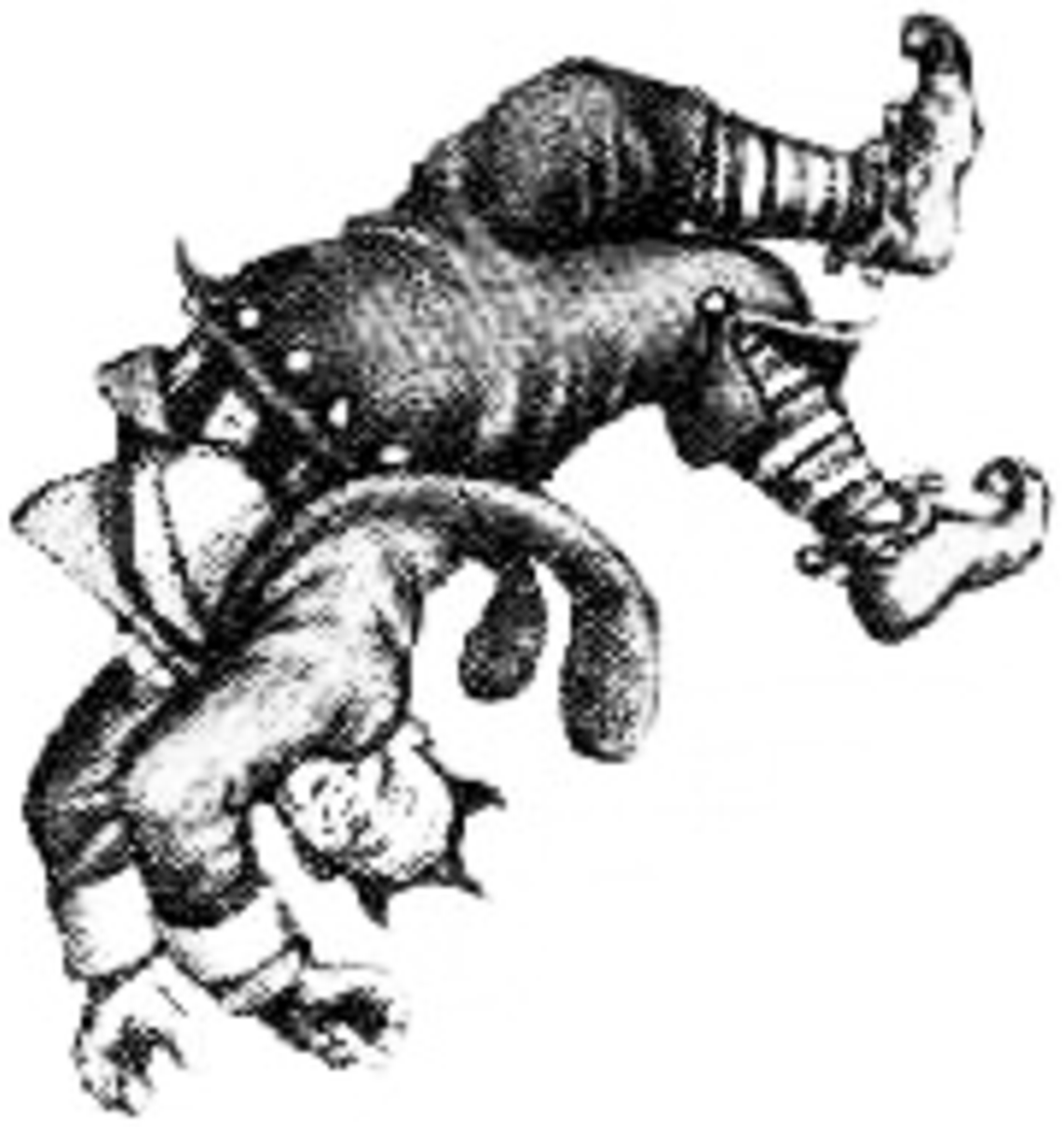}}
\end{figure}

Anyone stumbling onto M-Theory by chance would experience 
an overwhelming feeling that they are certainly 'not in Kansas anymore'. 
It takes some getting used to, the idea that we live in 11 dimensions;
I suspect it has a lot to do with a slight resentment that all our lives, 
so much has been going on that we were unaware of, but I'll leave that 
analysis for my dissertation in Psychology! For now, the point is to realise
that while it might seem a little overpowering initially, when Munchkins in
colourful costumes jump at you from all angles, 'Over the Rainbow' is
a wonderful place to be. 

\section{The Rainbow}

String theories have been around long enough that we are now more or less 
used to them. They are beautiful, but to the despair of many a string 
theorist a decade ago, there are almost as many consistent theories as 
colours in the rainbow! The theories looked lovely and were enjoyed
for their 'decorative' appeal but for a while it seemed that the
promise of unification which had resulted in so much excitement was 
not a promise that would be kept. 

Hope dawned anew in the mid-nineties when dualities were discovered 
and it was found that the string theories merely appear to be
different but in fact are all linked! We can move around from one to 
the other in a continuous circle and no one theory is any more
fundamental than the rest. So it was conjectured that perhaps there 
is something deeper, something which draws all the string
theories into its folds and unifies them. This something was given
the name M-Theory \cite{MTheoryreview}. 

In the history of string theory, we stand now at the moment of the
prism; all evidence points to the fact that the rainbow of string 
theories results from a single ray of light which is diffracted into 
the multi-coloured richness we see by the prism we unwittingly put 
in its path. However, for the time being, the white light of 
M-Theory is so bright that we are almost blinded.  
It will take a while before our eyes get accustomed to the light and
we can see our way clearly in this new land over the rainbow. 

\section{um ... M?}

Irreducible representations of a SUSY algebra dictate the possible
particle content of the corresponding supersymmetric field theory.
States in massless irreducible representations are labelled by helicity.
Using the SUSY generators, we can define fermionic creation
and annihilation operators which raise or lower helicity. Given the number
of supersymmetry generators $\cal N$ in the algebra and the maximal
helicity $\Lambda$ we want in a particular theory, we can then work out its
possible particle content.

Supergravity \cite{supergravity} is a theory of massless particles 
with helicity
$\Lambda \leq 2$. Maximally extended supergravity in 4 dimensions has
$8$ supersymmetry generators. Since these generators are Majorana spinors,
they have 4 components (supercharges) each which means that the maximum
number of supercharges in a supergravity theory is 32.
It takes all these supercharges to form a single 11-dimensional Majorana
spinor. Consequently, $D = 11$ is the highest
dimension in which a supergravity theory can exist.\\

\no
\underline{\bf \sf 11-d Supergravity}\\

The particle content of 11-dimensional supergravity \cite{11d} can be obtained
by studying irreducible representations of the algebra:
\be
\{Q_{\alpha}, Q_{\beta}\} = (\gamma^m C^{-1})_{\alpha \beta} P_m
+ (\Gamma^{mn} C^{-1})_{\alpha \beta} Z_{mn} +
(\Gamma^{mnpqr} C^{-1})_{\alpha \beta} Z_{mnpqr}
\ee
where $Q_{\alpha}$ is a Majorana spinor, $P_m$ is the momentum operator
and the $Z's$ are central charges. We find that the spectrum contains a 
graviton $G_{MN}$, a
rank three anti-symmetric tensor $A_{MNP}$ and a gravitino
$\psi^M_{\alpha}$.
The existence of all these fields can be intuitively justified: the
graviton
is needed for invariance under local coordinate transformations and the
gravitino is needed for invariance under local supersymmetric
transformations. However, the gravitino has 128 fermionic degrees of
freedom
whereas the graviton has only 44 bosonic degrees of freedom, so an
additional
84 bosonic degrees of freedom are needed to make the counting come out
right\footnote{All the counting of the degrees of freedom is done on shell} 
-- these are provided by the 3 form $A_{MNP}$. 

Just as a point particle is an electric source for a gauge field (also known 
as a one-form) $A_{M}$
through the coupling $e \int A_{M} dX^{M}$, a $p$-brane is an electric 
source for a $(p+1)$-form $A_{{M}_1 \dots {M}_{p+1}}$ through the coupling:
\be
\mu_{p} \int A_{{M}_1 \dots {M}_{p+1}} dX^{{M}_1} \wedge \dots \wedge 
dX^{{M}_{p+1}}
\ee
where $\mu_p$ is the charge of the $p$-brane under the
$(p+1)$-form. 
Moreover, a $p$-brane is also a magnetic source for the $(D - p - 3)$-form 
whose field strength is the Hodge dual in D dimensions 
of the $(p+2)$-form $F = dA$

We can hence see that the three-form $A_{MNP}$ in 11 dimensions 
couples electrically to a two-brane and magnetically to a five-brane. 
Accordingly, 11-dimensional supergravity has two types of branes, 
which are known as the M2-brane (or membrane) and the M5-brane or fivebrane.\\

\no
\underline{\bf \sf Dimensional Reduction}\\

Now consider what happens when the $X^{10}$ coordinate is a circle of
radius R. Since $X^{10} \equiv y$ is compact, we can Fourier expand 
11-dimensional fields as follows:
\be
\Phi_{11}(X) = \sum_n e^{\frac{iny}{R}} \Phi_{10}^n(X)
\ee
Each field $\Phi_{11}(X)$ in 11 dimensions leads to an infinite tower of
states $\Phi_{10}^n(X)$ in 10 dimensions, with masses that go like $n/R$. 
This can be seen 
by writing out the Klein-Gordon equation for the scalar field $\Phi$ 
\bea
\nabla^2_{11} \Phi_{11} = 0 \nonumber \\
\sum_n e^{\frac{iny}{R}} [\nabla^2_{10} - {\partial_y}^2] \Phi_{10}^n(X) = 0
\nonumber \\
\Rightarrow [\nabla^2_{10} - (n/R)^2] \Phi_{10}^n(X) = 0
\eea
At low energies or large distance scales, we do not see the eleventh direction.
From our ten-dimensional point of view, the momentum in this $y$ direction
hence seems like a mass. In the $R \rightarrow 0$ limit where the 
circle shrinks, all the states 
$\Phi_{10}^n(X)$ for $n \neq 0$ become infinitely massive so we
are left with a ten dimensional theory in which only the massless zero mode
$\Phi_{10}^0(x)$ survives.\\

Concentrating on just the bosonic fields, we find:
\be
\begin{array}[h]{c c c c c c c c}
\; & \nearrow & G_{\mu \nu} & \; & \; & \; & \nearrow &
B_{\mu \nu} = A_{\mu \nu 10}\\
G_{MN} & \rightarrow & A_{\mu} = G_{\mu 10} & \; & \; & A_{MNP} & \; & \;
\\
\; & \searrow & \Phi = G_{10,10} & \; & \; & \; & \searrow &
A_{\mu \nu \lambda}\\
\end{array}
\ee
The 11-dimensional metric gives rise to a metric $G_{\mu \nu}$,
a gauge field $A_{\mu}$ and a dilaton $\Phi$ in ten dimensions,
whereas the 3-form in eleven dimensions reduces to a 2-form
$B_{\mu \nu}$ and a 3-form $A_{\mu \nu \lambda}$ in ten
dimensions.\\

\no
\underline{\bf \sf IIA Supergravity}\\

In ten dimensions, we can arrange 32 supercharges to 
form two Majorana-Weyl spinors. If these spinors have the same 
chirality we get IIB supergravity 
and if they have opposite chiralities, we end up with IIA supergravity.
The particle content of IIA supergravity can be obtained by looking at
irreducible representations of the following algebra:
\bea
\{Q_{\alpha}, Q_{\beta}\} &=& (\Gamma^m C^{-1})_{\alpha \beta} P_m
+ (\Gamma^{mn} C^{-1})_{\alpha \beta} Z_{mn}  \\\nonumber
&+& (\Gamma^{mnpqr} C^{-1})_{\alpha \beta} Z_{mnpqr}
+ (\Gamma_{11} C^{-1})_{\alpha \beta} Z \\\nonumber
&+& (\Gamma^{m} \Gamma_{11} C^{-1})_{\alpha \beta} Z_{m} +
(\Gamma^{mnpq} \Gamma_{11} C^{-1})_{\alpha \beta} Z_{mnpq}\nonumber
\eea
The bosonic fields in this theory are a dilaton $\phi$, a metric
$G_{\mu \nu}$ a NS-NS 2-form $B_{\mu \nu}$, and R-R one and  three forms, 
$C_1$ and $C_3$.
Notice that this is precisely the spectrum obtained in the
$R \rightarrow 0$ limit of dimensional reduction of
11-dimensional supergravity on a circle.

\section{aheM!}

We have seen that IIA supergravity is a supersymmetric field
theory in its own right. However, it can also be
obtained as the low energy effective action of Type IIA string theory!
The $\alpha'\rightarrow 0$ or low energy limit is a consistent
truncation of string theory in which all massive states become
infinitely massive (since $M^2 \sim 1/{\alpha'}$) and hence decouple,
leaving behind only massless fields. The massless spectrum of
IIA string theory coincides exactly with the field content of IIA
supergravity.

Recall from the previous section that IIA supergravity can be obtained
from dimensional reduction of 11-dimensional supergravity compactified 
on a circle. More precisely, we have the following expression:
\be
ds^2_{11} = e^{- \frac{2}{3} \phi} ds^2_{10} +  e^{\frac{4}{3} \phi}
(dy^2 + A_{\mu} dX^{\mu}) 
\ee
Once again $y$ denotes the direction which 
has been compactified and $\phi$ is the dilaton. 
It can immediately be seen from the metric that
the radius R of the circular direction $y$ is given by 
\be
R = e^{\frac{2}{3} \phi} = g_s^{\frac{2}{3}} 
\nonumber
\ee
where we have used the fact that the dilaton is related
to the string coupling constant as $e^{\phi} = g_s$. Hence, 
small R corresponds to weak string coupling and large R to strong coupling. 

Putting all these facts together, we 
are lead to postulate the existence of a theory which would fill the 
missing corner in this rectangle of relations -- we call this M-Theory!

\begin{figure}[ht]
\epsfxsize=11cm
\centerline{\epsfbox{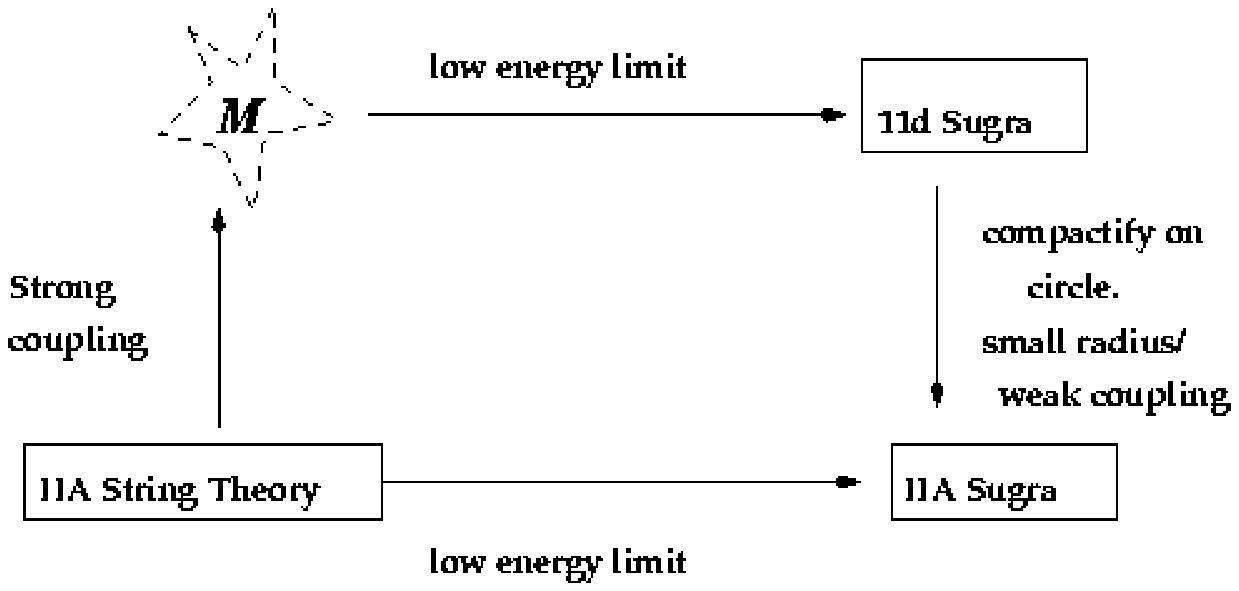}}
\end{figure}

Though we do not know yet what this theory is, we can deduce from the
above discussion what it reduces to in certain limits. The low energy limit of
M-Theory is 11d supergravity and the small radius limit
of M-Theory compactified on $S^1$ is IIA string theory! This is sometimes
turned around and stated as follows: The strong coupling limit of IIA
approaches a Lorentz invariant theory in 11 dimensions (M-Theory),
whose low energy limit is 11d supergravity.\\

\no
\underline{\bf \sf Branes descending to IIA}\\

We can have two types of M-branes, those transverse to $S^1$ and
those wrapped on it. An M$p$-brane wrapped on the circle 'loses' a
worldvolume direction when the circle is shrunk ($R \rightarrow 0$)
and has only $p-1$ spatial directions remaining when it arrives in
IIA. On the other hand, the spatial
extension of an M-brane transverse to the circle will not be
affected by the $R \rightarrow 0$ limit; such a brane will continue to have
a $p + 1$ dimensional world-volume in IIA. 

An M5-brane wrapped on the M-theory circle thus appears as a D4-brane
in Type IIA, whereas an M2-brane wrapped on 
the circle would become a
fundamental string. M5 and M2-branes transverse to the $S^1$ reduce to
NS5 and D2 branes respectively in IIA string theory \cite{dbranesfromm}. \\

\no
\underline{\bf \sf Supergravity as a detective.}\\

In what follows we will be dealing exclusively with BPS states as
solutions of 11-dimensional supergravity. The term BPS refers to the 
fact that these states saturate a bound which relates their mass to 
their charge $M \geq Q$. Charge conservation prevents the decay of
the least massive state with charge Q; it 
is obvious that the least massive state is the one which saturates the 
bound! It is also useful to note that supersymmetric states 
are automatically BPS.  

The wonderful thing about BPS states is that since they
are stable, they are expected to survive the perilous passage from 
supergravity to the full quantum theory of which the supergravity is
just a low energy limit. In other words, supersymmetric M-brane
configurations are present no matter how where we slide along 
the energy scale in
11 dimensions, moving from the all encompassing M-Theory to a small
sector therein. Since this small sector, supergravity, is currently
the only handle we have on M-Theory, we study M-branes in
this context and hope that the clues we find will help us build a 
picture of the elusive M-Theory. 

\section{Building Blocks}

Flat M2-branes and M5-branes are supersymmetric objects in M-Theory 
and can in fact be thought of as building blocks for the BPS spectrum since 
a large number of supersymmetric states can be constructed from them 
\cite{intbranes}. 

BPS states arise for example, when flat M-branes are wrapped on 
supersymmetric cycles, so-called because branes wrapped on them preserve
some spacetime supersymmetry. There are various other ways in which flat
branes can be combined to create supersymmetric configurations; a little 
later on, we will discuss the rule which dictates how these branes
must be placed such that the resulting set-up is BPS.   

We start by reviewing a few basic facts about M-branes.
In the expressions which follow, 
$X^{\mu}$ denotes coordinates tangent to the brane, $X^{\alpha}$ is
used to denote transverse coordinates and $r = \sqrt{X^{\alpha} X_{\alpha}} $
is the radial coordinate in this transverse space.\\

\no
\underline{\bf \sf The M5-brane:}\\

A flat M5-brane with worldvolume
$X^{{\mu}_0} \dots X^{{\mu}_5}$ is
a half-BPS object which preserves 16 real supersymmetries corresponding
to the components of a spinor $\chi$ which satisfies the condition:
\be
{\hat \Gamma}_{{\mu}_0{\mu}_1{\mu}_2{\mu}_3{\mu}_4{\mu}_5} \chi = \chi.
\ee
When this brane is placed in a flat background, the geometry is modified 
and the resulting spacetime is described by the following metric 
\bea
ds^2 &=& H^{-1/3} \eta_{\mu \nu} dX^{\mu} dX^{\nu} +
H^{2/3} \delta_{\alpha \beta} dX^{\alpha} dX^{\beta} \;\;\;\;\;\;\;\;\;\;\;\;\\
{\rm where}\;\;\;\;\;\;\;\;\;\;\;\;\;\;\;\;\;\;\; H &=& 1 + \frac{a}{r^3}
\label{flatm5metric}
\eea
and the four-form field strength 
\be
F_{\alpha \beta \gamma \delta} =
\frac{1}{2} \epsilon_{\alpha \beta \gamma \delta \rho}
\partial_{\rho} H.
\label{flatm54form}
\ee
of the supergravity three-form. 

Together, these equations (\ref{flatm5metric}) and (\ref{flatm54form}) 
specify the full
bosonic content of the supergravity solution for the M5-brane.\\

\newpage

\no
\underline{\bf \sf The M2-brane:}\\ 

A flat M2-brane spanning directions 
$X^{{\mu}_0}X^{{\mu}_1}
X^{{\mu}_2}$ is also a half-BPS object. Preserved spacetime supersymmetries
correspond to the 16 components of a spinor $\chi$ which survive the 
following projection
\be
{\hat \Gamma}_{{\mu}_0{\mu}_1{\mu}_2} \chi = \chi.
\ee
Like the M5 brane discussed above, the M2 brane is also a charged massive 
object which warps the flat space-time in which it is placed. 
The bosonic fields in the 
M2-brane supergravity solution are given by
\bea
ds^2 &=& H^{-2/3} \eta_{\mu \nu} dX^{\mu} dX^{\nu} +
H^{1/3} \delta_{\alpha \beta} dX^{\alpha} dX^{\beta} \;\;\;\;\;\;\;\;\;\;\;\;\\
F_{{\mu}_0{\mu}_1{\mu}_2 \alpha} &=&
\frac{ \partial_{\alpha} H}{2 H^2}, \;\;\;\;\;\;\;\;\;\;\;\;\\
{\rm where}\;\;\;\;\;\;\;\;\;\;\;\;\;\;\;\;\;\;\; H &=& 1 + \frac{a}{r^6}.
\eea

We now pause for a minute to discuss certain features of the landscape
which will guide us later when we try to navigate the 
supergravity solutions of more complicated brane configurations. \\

\no
\underline{\bf \sf Construction Site Rules}\\

One way of generating BPS states from the flat M-branes described above 
is to construct configurations of intersecting branes. In order for two 
M-branes 
to have a dynamic intersection, there must exist a worldvolume field to which 
this intersection can couple, either electrically or magnetically. 

Consider a $p$-brane which has a $q$-dimensional intersection with another
$p$-brane. From the point of view of the worldvolume, this
intersection must couple to a $(q + 1)$-form in order to be a 
dynamical object in the $(p+1)$
dimensional theory.

All $p$-branes contain scalar fields $\phi$ which describe their transverse
motion. The 1-form field strength of these scalars $F_1$ 
is the Hodge dual (on the worldvolume) of the $p$-form field 
strength $F_p$ of a $(p-1)$ form gauge field 
$A_{p-1}$, i.e
$$ d \phi = F_1 = * F_{p} = * d A_{p-1}$$
Since the gauge field $A_{p-1}$ couples to an object with 
$(p-2)$ spatial directions, we see 
that a $p$-brane can have a $(p-2)$ dimensional dynamical self intersection
\cite{p-2rule} .

This rule can be derived also using the BPS $\rightarrow$ no-force argument. 
Orienting branes of the same type so that they exert no force on each other, as
must be the case for stable supersymmetric configurations 
(in the absence of world-volume fields) it is found
that each pair of branes must share $(p-2)$ spatial directions.\\

\no
\underline{\bf \sf Intersecting Branes}\\

BPS states can thus be built from multiple M-branes if these are 
oriented such that each pair of M$p$-branes has a $(p-2)$-dimensional 
spatial intersection. Killing spinors of the resulting intersecting 
brane system are those which survive the projection 
conditions imposed by each of its flat M-brane constituents \cite{intbranes} .

Consider for example a system of multiple M2-branes. In order to obey the 
self intersection
rule, these membranes must be oriented such that no two membranes share 
any spatial 
directions. This criterion is satisfied by, for example, a system of four 
M2-branes 
with worldvolumes $012$, $034$, $056$ and $078$. The Killing spinors 
of this configuration are proportional to a
constant spinor $\eta$ obeying the following constraints:
\bea
{\hat \Gamma}_{012} \eta &=& \eta \\ \nonumber
{\hat \Gamma}_{034} \eta &=& \eta \\ \nonumber
{\hat \Gamma}_{056} \eta &=& \eta \\ \nonumber
{\hat \Gamma}_{078} \eta &=& \eta \nonumber
\eea
Since all the Gamma matrices\footnote{By Gamma matrices I mean the 
${\hat \Gamma}_{012}$ etc appearing in the 
above expression} here commute, they are simultaneously 
diagonalisable 
and we can proceed to search for eigenstates of the system. 

To begin with, notice that every Gamma matrix above squares to one so all 
the eigenvalues must be either 1 or -1. Since the trace of a matrix is the
sum of its eigenvalues, we know that each Gamma matrix has an equal
number of $\pm$ 1 eigenvalues, as all of the above matrices are
traceless. Using these matrices we can construct projection
operators\footnote{It is obvious that $P^+_i$ and $P^-_i$ are
projection operators as they obey
$$ (P^+_i)^2 = P^+_i, \;\;\;\; (P^-_i)^2 = P^-_i, \;\;\;\;  P^+_i + P^-_i = 1,
\;\;\;\; P^+_i P^-_i = 0 $$} for each Gamma matrix $\Gamma_i$ as follows: 
\be
P^+_i = \frac{1}{2} [1 + \Gamma_i] \;\;\;\;\;\; P^-_i = \frac{1}{2} 
[1 - \Gamma_i]
\ee
 
Acting $P^+_i$ on a spinor $\eta$, we see that components for which the
$\Gamma_i$ eigenvalue is -1 are projected out and the ones which
survive must have eigenvalue +1; thus the oft-quoted statement that Gamma
matrices (which square to one and are traceless) project out exactly
half the spinors. Acting now a second projection operator $P^+_j$, 
corresponding to another one of the Gamma matrices, we find the
following:
\be
P^+_j P^+_j \eta = \frac{1}{4} [1 + \Gamma_i + \Gamma_j +
\Gamma_i\Gamma_j] \eta 
\ee
Using the fact that the products of the Gamma matrices are also
traceless and square to one, we know that the eigenvalues of
$\Gamma_i\Gamma_j$ are also +1 and -1, in equal numbers. We now 
have the following 4 options corresponding to the possible
eigenvalues of $\Gamma_i$ and $\Gamma_j$:
\bit
\item
Both eigenvalues are -1 $\Rightarrow$ The spinor is projected out.
\item
The eigenvalues are +1 and -1 (or -1 and +1) respectively \\
$\Rightarrow$ 
Again, the spinor is projected out.
\item
Both eigenvalues are +1 $\Rightarrow$ The spinor survives.
\eit
Hence, only 1/4 of the spinors survive the combined projections due to
two Gamma matrices.

Extending this construction, it is easy to see that for a set of
$n$ independent\footnote{By which we mean that each matrix squares to one, as
do products of the matrices; each matrix is traceless, as are the
product matrices, and further all matrices in the set commute} Gamma
matrices the corresponding projection operators will project out an 
independent half of the supercharges, leaving behind $1/2^n$ supersymmetry.

Applying this to the intersecting membrane configuration described
above we see that the four projections imposed together leave behind 
only two supercharges, or equally, the 
brane configuration preserves $1/16$ supersymmetry.\\ 

\no
\underline{\bf \sf When Wrappings Become Intersections..}\\

A system of self-intersecting M$p$-branes corresponds to the singular limit of 
a single M$p$-brane wrapping a particular kind of smooth cycle. The cycle in 
question must be described by embedding functions that factorise. Each of the 
factors then gives the world-volume of a constituent brane. 

In order to clarify this somewhat complicated statement, consider the 
following 
simple example of a membrane wrapped on a holomorphic curve. Take this curve 
to be $f(u,v)= uv - c = 0$ in $\C^2$ where $c$ is a constant. In the limiting 
case when $c$ = 0 the curve becomes singular and the function $f$ obviously 
factorises to describe a system of two membranes spanning the $u$ and 
$v$-planes
respectively and intersecting only at a point. 

In general, a complex structure can be defined on the relative transverse 
directions (those which are common to at least one but not all of the 
constituents) of a system of intersecting branes. The intersecting brane 
configuration then describes the singular limit of an M-brane wrapping a 
smooth cycle embedded in this complex subspace of spacetime.     
Due to the $(p-2)$ self intersection rule, a system of $n$  
orthogonally intersecting membranes has a relative transverse space $\C^n$. 

Since preserved supersymmetries should be invariant under changes of
the constant $c$ in the holomorphic function $f(u,v)$, the Killing spinors 
of the wrapped brane configuration ($c \neq 0$) are the same as those 
for a system of $n$ orthogonal membranes (the $c = 0$ limit) intersecting 
according to the $(p-2)$ rule. 

\section{Landmarks}

The Munchkins have now brought you to the end of their domain. 
To get to your destination though, you'll just have to 
'Follow the Yellow Brick Road'. However, during your long walk, 
it might be useful to keep in mind some of the things you have 
learnt in this new land Over The Rainbow

On the world-volume of a flat M-brane, we would expect to have Poincare 
invariance. Hence, a metric describing space-time in the presence of 
such a brane should not have any dependence on coordinates tangent 
to the M-brane. Also, 
we expect the 
effects of the brane on spacetime to decrease as we move away from it and 
furthermore, the 
configuration is invariant under rotations in the transverse directions.
 From these considerations
of isometry alone, we can conclude that the metric depends only on 
the radial coordinate
in the space transverse to the flat M-brane world-volume. 
Notice that this is true 
for both the 
M-brane solutions discussed previously. 

Moreover, each solution can be specified completely in terms of not 
just an arbitrary but in fact a {\bf harmonic} function of 
this radial coordinate. Tracing the origin of this condition, we are lead 
to the 
equation of motion for the field strength $d * F = 0$. Writing this out in 
component form,
we find 
\be
\partial_{I} (\sqrt{|{\rm det} g_{IJ}|} F^{IJKL}) = 0 
\ee
which implies that 
\be
{\nabla}^2 H = \Sigma_{\alpha} \; \frac{{\partial}^2 H}{{{\partial} 
X_{\alpha}}^2} = 0
\ee
so the function H must be a solution to the flat space Laplacian. Both 
these characteristics (i.e. the isometries and the role of the harmonic 
function) 
will have interesting generalisations when we consider 
supergravity solutions for non-trivial brane configurations. \\

\chapter{The Yellow Brick Road}

\begin{figure}[ht]
\epsfxsize=8cm
\centerline{\epsfbox{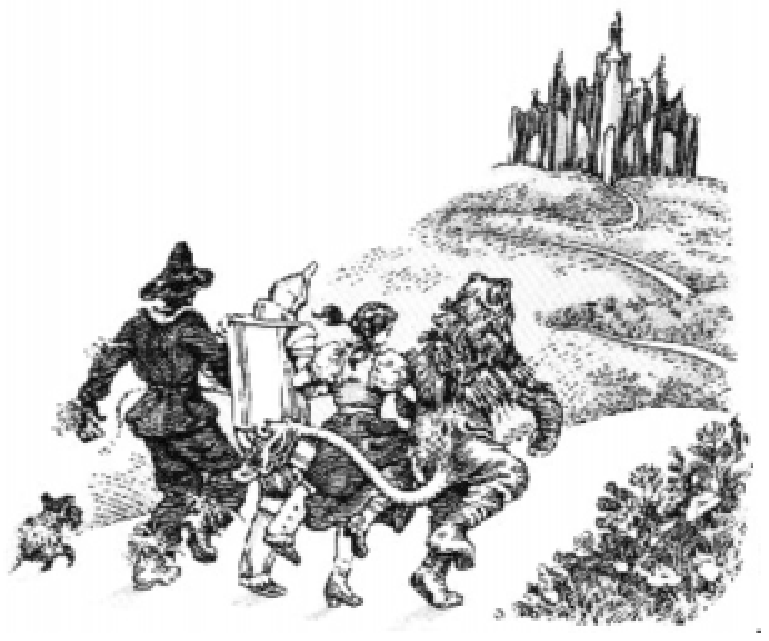}}
\end{figure}

In the paper \cite{harmfn} where the Harmonic Function rule was proposed, 
before
even stating what the rule was, 
Tseytlin pointed out the assumptions that went into it. We should bear these in
mind so that we only apply the rule to the configurations it was designed to 
describe 
in the first place. For starters, 
\bit
\item
The harmonic function rule can be used only to construct
supergravity solutions for intersecting $p$-brane systems which are 
smeared along the relative transverse directions (those which 
are tangent to atleast one but not all of the constituent branes).
 As a result, the metric is then independent
of these coordinates and is a function only of the 
overall transverse directions, which are not tangent to 
the world-volume of any brane in the system.
\eit
It is mentioned in passing that a class of 
more general solutions is expected to exist, such that each constituent 
brane 
has a different transverse space. As we will see later on, these are the 
cases covered by the Fayyazuddin-Smith metric ansatz! 
\bit
\item
Secondly, since these supergravity solutions are governed by harmonic 
functions of
the radial coordinate in the transverse space, this (overall) 
transverse space must be at least 
three dimensional if the harmonic functions are to decay at infinity. 
\eit
This excludes from the present consideration certain intersecting M-brane 
systems which
{\bf are} allowed by the $(p-2)$ self-intersection rule \cite{p-2rule}. 
As will later be seen, 
these missing configurations are encompassed by the Fayyazuddin-Smith ansatz. 
\bit
\item
Lastly, it is useful also to remember that the only configurations to which 
the rule is
expected to be applicable are those for which the Chern-Simons contribution 
to the 
equation of motion for the four-form vanishes, i.e $F \wedge F = 0.$ 
\eit

Bosonic backgrounds constructed by combining the individual supergravity 
solutions
of constituent branes can still be solutions to the equations of motion of 
D=11 supergravity.
The basic observation is that for brane bound states with
zero binding energy, it is possible to assign an independent harmonic
function to each constituent intersecting brane. The argument for this is 
sketched below.

A single brane supergravity solution can be expressed completely in terms of 
one harmonic function. For an extremal (no binding energy) BPS configuration 
of N branes, there is no force between the constituents and hence no
obstruction to moving one of the branes far apart from the
others. When a brane is moved sufficiently far apart from the rest,
their effects on it are negligible and it is to all intents 
and purposes free. Fields near it should thus approximate the fields 
in the supergravity solution of a single brane. Since any or all N of the 
branes can arbitrarily be moved back and forth at no cost to the
energy, we would expect the solution describing a configuration of N
branes to be parameterized by N independent harmonic functions. 

\section{The Rule of the Harmonic Functions}

In the presence of an intersecting M-brane configuration, the $(10+1)$ 
dimensional spacetime naturally 'splits up' into three seperate
parts.
The directions common to all M-branes are referred to as
the {\bf common tangent directions}. Tangent space indices here 
will be denoted by $a,b$, and curved space indices by $\mu \nu$. 
In the wrapped brane picture, these are the worldvolume directions
which are left flat.
The supersymmetric cycle which the M-brane wraps is embedded 
in to the subspace spanned by the {\bf relative transverse directions}, 
so called as they are tangent to at least one but not all of the 
constituent branes. We define complex coordinates on this space where 
flat and curved indices are denoted by 
$m,\overline{n}$, and
$M,\overline{N}$ respectively.
Finally there are the {\bf overall transverse directions} which span 
the space transverse to all constituent branes in the intersecting
brane system; in the wrapped brane picture, these are the directions 
which are transverse to both
the brane worldvolume and the embedding space. 

\subsection*{It is Decreed $\cdots$}

As mentioned above, the harmonic function rule gives a recipe for 
'superposing' the
individual bosonic fields in the supergravity solutions of each of the
component branes in an intersecting brane system.
\bit
\item
{\bf \sf \underline{The Metric}}\\
Assigning a harmonic function $H_a$ to each constituent
M-brane, we proceed to construct the metric for a multi-brane
configuration by taking our cue from the metric for a single brane. 
In analogy to that case, directions tangent to the
$i^{th}$ M5-brane are multiplied by a factor of $H_i^{-1/3}$ whereas 
directions transverse to it are multiplied by $H_i^{2/3}$.

Similarly, a metric describing the background created by a system
of intersecting membranes, can be constructed by ensuring that 
the coordinates along the worldvolume of the $j^{th}$ M2-brane
carry a factor of $H_j^{-2/3}$ while transverse coordinates are multiplied by 
$H_j^{1/3}$.
\item
{\bf \sf \underline{The Field Strength}}\\

Since the field strength components due to each constituent M-brane 
carry different indices, the field strength of the intersecting brane
configuration can be obtained merely by adding the individual field 
strengths corresponding to each M-brane.
\eit

\no
This rule is made clear by its application to the following systems. 

\subsection*{Two Membranes}

\be
\begin{array}[h]{|c|c|cccc|cccccc|}
  \hline
   \; & 0 & 1 & 2 & 3 & 
              4 & 5 & 6 & 7 & 8 & 9 & 10\\
  \hline
  {\bf M2} & \times & \times & \times &  &  &  
               &  &  &  &  &  \\
  {\bf M2} & \times &  &  & \times
               & \times &  &  &  &  &  &  \\
  \hline
\end{array}
\ee
Assigning a factor of $H_1$ to the first membrane and $H_2$ to the second, 
we use
the harmonic function rule to write down the metric and field
strength. The metric is given by
\bea
ds^2 &=& H_1^{1/3}  H_2^{1/3} [ - H_1^{-1} H_2^{-1} dX_{0}^2 \nonumber 
\\
&+&  H_1^{-1} (dX_{1}^2 + dX_{2}^2) +  H_2^{-1} (dX_{3}^2 + dX_{4}^2) 
\nonumber 
\\
&+& (dX_{5}^2 + dX_{6}^2 + dX_{7}^2+ dX_{8}^2 + dX_{9}^2 +dX_{10}^2)]
\eea
and the non-vanishing components of the field strength are
\be
F_{012\alpha} = \frac{\partial_{\alpha} H_1 }{H_1^2} \; \; \; \; 
F_{034\alpha} = \frac{\partial_{\alpha} H_2 }{H_2^2}
\ee
where $H_1$ and $H_2$ are functions only of  $X^{\alpha}$ for $\alpha
= 5, 6 \dots 10$.\\

\subsection*{Three Membranes}

Introducing now a third membrane, characterized by the harmonic function 
$H_3$, we have 
the following configuration
\bea
\begin{array}[h]{|c|c|cccccc|cccc|}
  \hline
   \; & 0 & 1 & 2 & 3 & 
              4 & 5 & 6 & 7 & 8 & 9 & 10\\
  \hline
  {\bf M2} & \times & \times & \times &  &  &  
               &  &  &  &  &  \\
  {\bf M2} & \times &  &  & \times
               & \times &  &  &  &  &  &  \\
{\bf M2}  & \times &  &   &  &  & \times
               & \times &  &  &  &  \\
  \hline
\end{array}
\eea
with metric
\bea
ds^2 &=& H_1^{1/3}  H_2^{1/3} H_3^{1/3} [- H_1^{-1} H_2^{-1} H_3^{-1} 
dX_{0}^2 +  H_1^{-1} (dX_{1}^2 + dX_{2}^2) \nonumber \\
&+& H_2^{-1} (dX_{3}^2 + dX_{4}^2) +  H_3^{-1} (dX_{5}^2 + dX_{6}^2) 
\nonumber \\
&+& dX_{7}^2+ dX_{8}^2 + dX_{9}^2 +dX_{10}^2]
\eea
and field strength,
\be
F_{012\alpha} = \frac{\partial_{\alpha} H_1 }{H_1^2} \;\;\;\;
F_{034\alpha} = \frac{\partial_{\alpha} H_2 }{H_2^2} \;\;\;\;
F_{056\alpha} = \frac{\partial_{\alpha} H_3 }{H_3^2} 
\ee
Here, $H_1$, $H_2$ and $H_3$ are functions of the transverse
directions $X^{\alpha}$ where $\alpha = 7 \dots 10$

\subsection*{Two Fivebranes}

\be
\begin{array}[h]{|c|cccc|cccc|ccc|}
  \hline
   \; & 0 & 1 & 2 & 3 & 
              4 & 5 & 6 & 7 & 8 & 9 & 10\\
  \hline
  {\bf M5} & \times & \times & \times & \times & \times & \times
               &  &  &  &  &  \\
  {\bf M5} & \times & \times & \times & \times
               &  &  & \times & \times &  &  &  \\
  \hline
\end{array}
\ee
The harmonic function rule dictates the following metric
\bea
ds^2 &=& H_1^{2/3}  H_2^{2/3} [H_1^{-1}  H_2^{-1} 
(- dX_{0}^2 + dX_{1}^2 + dX_{2}^2 + dX_{3}^2) 
\nonumber \\
&+&  H_1^{-1} (dX_{4}^2 + dX_{5}^2) + H_2^{-1} (dX_{6}^2 + dX_{7}^2) 
\nonumber \\
&+& (dX_{8}^2 + dX_{9}^2 +dX_{10}^2)]
\eea
and field strength components
\bea
F_{67 \alpha \beta} = \epsilon_{\alpha \beta \gamma} H_1 \nonumber \\
F_{45 \alpha \beta} = \epsilon_{\alpha \beta \gamma} H_2
\eea
The functions $H_1$ and $H_2$ depend only on the overall transverse
directions labelled by $\alpha$ which takes values $8,9,10.$\\

\subsection*{Three Fivebranes}

Adding now a third M5-brane in the following manner:
\be
\begin{array}[h]{|c|cc|cccccc|ccc|}
  \hline
   \; & 0 & 1 & 2 & 3 & 
              4 & 5 & 6 & 7 & 8 & 9 & 10\\
  \hline
  {\bf M5} & \times & \times & \times & \times & \times & \times
               &  &  &  &  &  \\
  {\bf M5} & \times & \times & \times & \times
               &  &  & \times & \times &  &  &  \\
{\bf M5} & \times & \times &  &  
               & \times & \times & \times & \times &  &  &  \\
  \hline
\end{array}
\ee
we find that spacetime is described by the metric
\bea
ds^2 &=& H_1^{2/3}  H_2^{2/3} H_3^{2/3}
[H_1^{-1}  H_2^{-1} H_3^{-1}(- dX_{0}^2 + dX_{1}^2) \nonumber \\
&+& H_1^{-1}  H_2^{-1} (dX_{2}^2 + dX_{3}^2)
+  H_1^{-1} H_3^{-1} (dX_{4}^2 + dX_{5}^2) \nonumber \\ 
&+& H_2^{-1} H_3^{-1} (dX_{6}^2 + dX_{7}^2)
+ (dX_{8}^2 + dX_{9}^2 +dX_{10}^2)]
\label{hfnm5c3}
\eea
and field strength components are
\bea
F_{67 \alpha \beta} = \epsilon_{\alpha \beta \gamma} H_1 \nonumber \\
F_{45 \alpha \beta} = \epsilon_{\alpha \beta \gamma} H_2 \nonumber \\
F_{23 \alpha \beta} = \epsilon_{\alpha \beta \gamma} H_3 \nonumber
\eea
\label{fstm5c3}
The harmonic functions $H_1$, $H_2$ and $H_3$ depend on $X^{\alpha}$
where $\alpha = 8,9,10$.

\section*{Exiled!}

Under the Rule of the Harmonic Functions, the configurations described
below are banished on the charge that they do not have a sufficient
number of overall transverse directions. The minimum number of
transverse directions required for a law abiding brane configuration
living under the Harmonic Function rule is three. At least three overall
transverse directions are needed in order for the functions $H_i$
to be solutions of the flat Laplacian in this subspace with the right 
behaviour at infinity and thus be deserving
of the title 'Harmonic' functions.\\ 

\no \underline{\bf \sf Two Transverse Directions}

\be
\begin{array}[h]{|c|c|cccccccc|cc|}
  \hline
   \; & 0 & 1 & 2 & 3 & 
              4 & 5 & 6 & 7 & 8 & 9 & 10\\
  \hline
  {\bf M2} & \times & \times & \times &  &   &  
               &   &   &   &   &   \\
  {\bf M2} & \times &   &   & \times
               & \times &   &   &   &   &   &  \\
{\bf M2}  & \times &   &    &   &   & \times
               & \times &   &   &   &  \\
{\bf M2} & \times &   &   &   &   &   &   & \times
               & \times  &   &  \\
 \hline
\end{array}
\ee

\no \underline{\bf \sf One Transverse Direction}

\be
\begin{array}[h]{|c|cccc|cccccc|c|}
  \hline
   \; & 0 & 1 & 2 & 3 & 
              4 & 5 & 6 & 7 & 8 & 9 & 10\\
  \hline
  {\bf M5} & \times & \times & \times & \times & \times & \times
               &  &  &  &  &  \\
  {\bf M5} & \times & \times & \times & \times
               &  &  & \times & \times &  &  &  \\
{\bf M5} & \times & \times & \times & \times
               &  &  &  &  & \times & \times &  \\
  \hline
\end{array}
\ee

\no \underline{\bf \sf No Transverse Directions!}

\be
\begin{array}[h]{|c|c|cccccccccc|}
\hline
   \; & 0 & 1 & 2 & 3 & 
              4 & 5 & 6 & 7 & 8 & 9 & 10\\
\hline
  {\bf M2} & \times & \times & \times &  &  &  
               &  &  &  &  &  \\
  {\bf M2} & \times &  &  & \times
               & \times &  &  &  &  &  &  \\
{\bf M2}  & \times &  &   &  &  & \times
               & \times &  &  &  &  \\
{\bf M2} & \times &  &  &  &  &  &  & \times
               & \times  &  &  \\
{\bf M2} & \times &  &  &  &  &  &  &  &  
	       &\times & \times  \\
\hline
\end{array}
\ee

\section{The Cycles we Turn To}

Supersymmetric cycles have the defining property that branes
wrapping them preserve some supersymmetry. Holomorphic cycles are known
to be supersymmetric; in fact being the simplest examples of
supersymmetric cycles they are the ones considered most often. All the 
configurations studied in this thesis describe M-branes
wrapping holomorphic cycles, so we now take a small detour and see
how holomorphicity leads to supersymmetry and study the
conditions it imposes on the surviving supercharges. 

A wrapped $p$-brane whose embedding into spacetime is described by 
$X^{\mu}(\sigma^i)$ is said to be supersymmetric only if the 
background it gives
rise to admits at least one Killing spinor $\chi$ such that \cite{BBS}
\be
\chi = \frac{1}{p!} \frac {1}{\sqrt h} {\hat{\epsilon}}^{\alpha_1 ... \alpha_p}
\Gamma_{M_1 ... M_p}
\partial_{\alpha_1} X^{M_1} ....
\partial_{\alpha_p} X^{M_p} \chi. 
\label{KSeqn}
\ee
where $\Gamma_{M_1 ... M_p}$ is the completely anti-symmetrized product
of $p$ eleven dimensional $\Gamma$ matrices,
$\hat{\epsilon}$ is the 
$p$-dimensional Levi-Civita symbol and $h$ denotes the
determinant of the induced metric $h_{ij}$ on the brane.

For a brane embedded in a complex space with Hermitean metric $G_{U \bar{V}}$,
the induced metric on the worldvolume is 
\be
h_{ij} = [\partial_i X^{U} \partial_j X^{\bar V} + 
\partial_j X^{U} \partial_i X^{\bar V}] G_{U \bar{V}}
\ee
If the brane is holomorphically embedded, the supersymmetric cycle it wraps 
must be even-dimensional; we denote its dimension by $2m$. 
Defining now a complex structure on this cycle which is compatible
with the complex structure in spacetime, the induced metric can be 
written in the form:
\be 
h_{u {\bar v}} = \partial_u X^{U} \partial_{\bar v} X^{\bar V} G_{U \bar{V}}
\ee
Hermiticity of the embedding space metric $G_{U \bar{V}}$, will imply
that the induced metric on the worldvolume is Hermitean as well; as
such ${\rm det} \; h_{u \bar{v}} = \sqrt{h}$ is given by
\be
\sqrt{h} = \frac{1}{m!} \epsilon^{u_1 \dots u_m} \epsilon^{\bar{v_1} \dots 
{\bar v_m}} 
h_{u_1 {\bar v_1}} \dots h_{u_m {\bar v_m}}
\label{deth}
\ee
The Killing spinor equation (\ref{KSeqn}) now takes the form \cite{AnsarMichal}
\bea
\sqrt{h} \chi &=& (\frac{1}{m!})^2 {\hat{\epsilon}}^{u_1 ... u_m}
\epsilon^{\bar{v_1} \dots {\bar v_m}} 
\Gamma_{U_1 \bar{V}_1... U_m \bar{V}_m}
\partial_{u_1} X^{U_1} 
... \partial_{{\bar v}_m}
X^{{\bar V}_m} \chi \nonumber \\
&=& (\frac{1}{m!})^2 {\hat{\epsilon}}^{u_1 ... u_m}
\epsilon^{\bar{v_1} \dots {\bar v_m}} 
\Gamma_{u_1 \bar{v}_1... u_m \bar{v}_m}
\eea
Inserting the expression for $h$ from (\ref{deth}), we can
read off the projection conditions on the Killing spinor associated
with an M-brane wrapping the given $2m$-dimensional holomorphic
cycle. 

\no
In particular, for a two-cycle, we find  
\be
\Gamma_{u \bar{v}} \chi = h_{u {\bar v}} \chi
\ee
and for a four-cycle,
\be
\Gamma_{u \bar{v} s {\bar s}} \chi = 
[h_{u {\bar v}} h_{w {\bar s}} - h_{u {\bar s}} h_{w {\bar v}}]  \chi
\ee
Both these conditions will be used later on when we discuss
holomorphic embeddings of M-branes. 

\bigskip
\begin{center}
\noindent\fbox {\noindent\parbox{4.5in}{{\sf \bf The Determinant of a 
Hermitean Metric.}\\

Assume that the metric $h_{M N}$ in an $m$ dimensional complex space is 
Hermitean. Hermiticity implies that $h_{i j} = h_{\bar{i j}} = 0$. Thus, 
in the $2m \times 2m$ dimensional matrix which would conventionally be used to 
represent the metric, there are only $4m^2 - m^2 - m^2$ entries which are 
non-zero and 
none of these are along the diagonal. Using then the fact that a metric 
must be symmetric, we find that the degrees of freedom are reduced by a 
further half, leaving only $m^2$ entries to determine the metric completely. 
These can be arranged into an $m \times m$ matrix which specifies 
the Hermitean metric.

Consider a simple example to illustrate this point: Let $h_{M N}$ be the metric
in a $2m$ real dimensional space. 
$$ {\rm det} \; h_{M N} = \frac{1}{(2m)!} \epsilon^{I_1 \dots I_{2m}} 
\epsilon^{J_1 \dots J_{2m}} h_{I_1 J_1} \dots h_{I_{2m} J_{2m}} $$
If a complex structure is defined on this space, the anti-symmetric
tensor $\epsilon^{I_1 \dots I_{2m}}$ splits up into a product of
two tensors, with holomorphic and anti-holomorphic indices as follows:
$$ \frac{1}{(2m)!} \epsilon^{I_1 \dots I_{2m}} = \frac{1}{m!} 
\epsilon^{i_1 \dots i_{m}} \frac{1}{m!} 
\epsilon^{\bar{j}_1 \dots \bar{j}_{m}} $$
If this complex structure is such that the resulting 
metric is hermitean, the determinant can be written as
$$
h \equiv {\rm det} \; h_{M N} = [{\rm det} \; h_{i {\bar{j}}} ]^2 $$
where 
$$ {\rm det} \; h_{i {\bar{j}}} =  \frac{1}{m!} \epsilon^{i_1 \dots
i_m} \epsilon^{{\bar{j}}_1 \dots {\bar{j}}_m} h_{i_1 {\bar{j}}_1} 
h_{i_2 {\bar{j}}_2}  \dots h_{i_m {\bar{j}}_m}$$}}
\end{center}
\bigskip

\section{Killing (some) Spinors}

The amount of supersymmetry preserved by a p-brane with
worldvolume $X^{M_1}...X^{M_p}$ is given by the number of spinors which
satisfy (\ref{KSeqn}).

Because we are dealing with holomorphic cycles, and a complex structure 
has been defined on the embedding space, it makes sense to re-write the 
Clifford algebra in complex coordinates as well. This is in fact a very 
useful thing to do, as it turns out that the flat space Clifford 
algebra when expressed in complex coordinates resembles the algebra of 
fermionic creation and annihilation operators! We can thus express 
spinors on a complex manifold as states in a Fock space \cite{GSW}. Defining
$\Gamma$ matrices for a complex coordinate $z_j = x_j + iy_j$ as follows:
\bea
\Gamma_{z_j} &=& \frac{1}{2} (\Gamma_{x_j} + i \Gamma_{y_j})\\ \nonumber
\Gamma_{\overline{z_j}} &=& \frac{1}{2} (\Gamma_{x_j} - i \Gamma_{y_j})
\eea
the Clifford algebra in $C^n$ takes the form
\be
\{ \Gamma_{z_i}, \Gamma_{\overline{z_j}} \} = 2 \eta_{i \bar{j}}.
\ee
Declaring $\Gamma_{z_i}$ to be creation operators and
$\Gamma_{\overline{z_j}}$ to be annihilation operators, it is now clear that
a Fock space can be generated by acting the creation operators on a vacuum. 
Because there are $n$ creation operators, each state in the
Fock space is labelled by $n$ integers taking values 0 or 1 which 
correspond to its fermionic occupation numbers.

We illustrate the utility of this construction by considering in detail an
M5-brane wrapping a two-cycle in $\C^3$. Let the holomorphic 
coordinates in $\C^3$ be $u,v,w$. The Clifford algebra can then be
explicitly written out as 
\be
\{ \Gamma_u, \Gamma_{\overline{u}} \} =
\{ \Gamma_v, \Gamma_{\overline{v}} \} =
\{ \Gamma_w, \Gamma_{\overline{w}} \} = 1;
\ee
all other anti-commutators are zero.
Since the Fock vacuum is the state where none of the oscillators
are excited it is denoted by $|000>$ and the highest weight state where
all oscillators are excited is denoted by $|111>$. Hence we have
\bea
\Gamma_{\overline{z}} |000> &=& 0 \\ \nonumber
\Gamma_{z} |111> &=& 0.
\eea
where z can take values $u, v,w$.
A generic spinor $\psi$ in $\C^3$ can then be decomposed in terms of Fock
space states as follows:
\bea
\psi &=& a |000>  + b |100>  + c |010> + d |001> \\\nonumber
     &+& e  |110> + f |101> + g |011> + h |111>
\eea

An eleven dimensional spinor $\chi$ can be written as a sum of terms of
the form $\alpha \otimes \psi \equiv \alpha \otimes |n_u, n_v, n_w>$ where 
$\alpha$ is
a four-dimensional spinor and $n_z ,\; {\rm for} \; z=u,v,w$ are the
fermionic occupation numbers of the state. Expressing spinors in this way
and writing down the supersymmetry preservation conditions in
terms of $\Gamma_z$, it becomes very easy to figure out the Killing spinors. 

Consider an M5-brane wrapping a two-cycle in $\C^3$. From (\ref{KSeqn}) 
we see that supersymmetry is preserved only if solutions can be found to the 
equation \cite{Us}
\bea
i \Gamma_{0123} \Gamma_{m \bar{n}} \chi = \eta_{m \bar{n}} \chi
\label{M5C3susycond}
\eea
For the time being, we restrict ourselves to the implications of the above 
condition on $\psi$ alone. We then find the following constraints:
\bea
\Gamma_{u \bar{v}} \psi = 
\Gamma_{u \bar{w}} \psi = 
\Gamma_{v \bar{u}} \psi = 0 \nonumber \\
\Gamma_{v \bar{w}} \psi = 
\Gamma_{w \bar{u}} \psi =  
\Gamma_{w \bar{v}} \psi = 0
\label{killspinors}
\eea

\bigskip
\begin{center}
\noindent\fbox {\noindent\parbox{4.5in}{{\sf \bf Those which were Killed, 
Live!}\\

We pause for a minute to see what these constraints say about the spinor. Take
a simple constraint, say $\Gamma_{u} \chi = 0$. 
Since $\Gamma_{u}$ is a creation operator, it acts on $\psi$ such that 
all states with $n_u = 0$ are taken to $n_u = 1$ but states which already 
had $n_u = 1$ are killed. Writing this out we find
$$\Gamma_{u} \psi = a |100> + b |110> + c |101> + d |111> = 0$$
Since all Fock space states are independent, this implies that
$a = b = c = d = 0$.
The spinor $\psi$ can now only have the components $|000>, |010> , |001>$ and 
$|011>$ 
Hence, in effect what a constraint of the type $\Gamma \psi = 0$ does is to
kill the spinors which survive the action of $\Gamma$ and keep the spinors
which were eliminated!}}
\end{center}
\bigskip 
\no
Hence (\ref{killspinors}) eliminates all states which have the following 
occupation numbers. 
\bea
n_u = 0 \; {\rm and} \; n_v = 1, \;\; {\rm or}  \;\; n_u = 0 \; {\rm and} \; 
n_w = 1, \nonumber \\ 
n_v = 0 \; {\rm and} \; n_u = 1, \;\; {\rm or} \;\; n_v = 0 \; {\rm and} \; 
n_w = 1, \nonumber \\ 
n_w = 0 \; {\rm and} \; n_u = 1, \;\; {\rm or} \;\; n_w = 0 \; {\rm and} \; 
n_v = 1. 
\eea
The only components of $\psi$ which survive this treatment are $|000>$ 
and $|111>$. Recall that (\ref{M5C3susycond}) also gives rise to constraints
\bea
i \Gamma_{0123} \Gamma_{z \bar{z}} \chi = \frac{1}{2} \chi 
\label{spinorchirality}
\eea 
Since 
\bea
\Gamma_{z \bar{z}} |000> &=& |000> \nonumber \\
\Gamma_{z \bar{z}} |111> &=& - |111>
\eea
these constraints can be used to determine the four-dimensional chirality of 
the spinors $\alpha$ and $\beta$ when $\chi$ is expressed as 
\be
\chi = \alpha \otimes |000> + \beta \otimes |111>
\ee
Imposing (\ref{spinorchirality}) we find that 
\bea
i \Gamma_{0123} \alpha &=& \alpha \nonumber \\
i \Gamma_{0123} \beta &=& - \beta 
\eea

\subsection*{Counting}

The last step is to count the number of supercharges preserved by this
configuration. As a generic spinor in 11 dimensions, $\chi$ starts out 
life with 32 complex components: 4 complex components come from 
the Dirac spinor $\alpha$ and, as we have seen
explicitly, 8 components come from $\psi$. After the constraints 
(\ref{killspinors}) are imposed, only 2 of these 8 components survive, 
so the spinor $\chi$ is left with $4 \times 2$ complex degrees of
freedom. Determining the chirality of $\alpha$ and $\beta$ cuts the 
degrees of freedom down by a further half. Finally we impose the 
Majorana condition, which essentially states that  $\chi$ can be 
completely determined by $\alpha |000>$ and thus has only the 4 real 
degrees of freedom of this state. 

An M5-brane wrapping a holomorphic 
two-cycle in $C^3$ hence preserves 4 of the 32 possible supercharges, or 
$1/8$ of the total spacetime supersymmetry.

\section{Pushing the Boundaries }

Now that we have a rule which allows us to contruct supergravity
solutions for a certain class of BPS bound states of branes, we can
try to extend the scope of this rule to a wider class of
supersymmetric configurations. A natural step in this direction
 would be to make the
harmonic functions depend on coordinates in the relative transverse
space so that they may describe branes which are not smeared in this
subspace. However, even if we start from this assumption it turns out
that at least one of the branes becomes delocalized because of the 
translational invariance which results from smearing.

To illustrate this point consider the simplest possible
example, that of 2 M2-branes, which span worldvolumes $X^0 X^1 X^2$ and 
$X^0 X^3 X^4$ intersecting at a point, and are characterized by 
functions $H_1$ and $H_2$ respectively.  
$H_1$ and $H_2$ are now allowed to depend on all spatial coordinates,
in contrast to what was acceptable previously for the harmonic
function rule. 

If the membrane spanning $X^0 X^3 X^4$ is localized then 
we should have translational invariance on its world-volume implying that the 
membrane with worldvolume $X^0 X^1 X^2$ cannot be
localized in $X^3$ or $X^4$ and must instead be smeared.  
So, as far as this M2-brane is concerned, life is just as it was 
previously under the harmonic function rule; $H_1$ still 
depends only on the overall transverse directions $X^5 \dots
X^{10}$: Moreover, since it also obeys the flat space Laplace 
equation in these directions, $H_1$ is a Harmonic function of 
the radial coordinate in this space. 

However, since the M2-brane with worldvolume $X^0 X^3 X^4$ is localised, the
function $H_2$ characterising it obeys a different equation. Recall that
the origin of the Harmonic function condition was 
the equation of motion for the four-form $d*F=0$. Applying this to the
case at hand, we now find that $H_2$ obeys the {\bf curved space} 
Laplace equation \cite{wrapM5semilocalised}
\be
H_2 ({\partial_3}^2 + {\partial_4}^2) H_1 + ({\partial_5}^2 \dots
{\partial_{10}}^2) H_1 = 0
\ee
and is hence called a {\bf generalised harmonic function} 

\section{The Kingdom is larger than it seems..}

The Harmonic Functions rule very successfully in their own domain, but
they never claimed to exercise absolute control over the entire 
kingdom of supergravity
solutions. Branes smeared over the relative transverse directions
are loyal subjects, (as long as they have three or more overall
transverse directions) but, as we have seen in the previous section, 
localised intersections defy the Rule and consequently
must be described by a wider class of spacetimes.

In an attempt to address this problem, Fayyazuddin and Smith \cite{FS1}
came up with a metric ansatz whose form is dictated by the 
isometries of the background in 
the presence of a wrapped brane configuration,(which could have
an intersecting brane system as its singular limit). 

The spacetimes 
they considered describe M-branes wrapping holomorphic cycles so a 
complex structure has been described on the embedding space in order to
make holomorphicity transparent. If the supersymmetric cycle 
is embedded into an $2n$ dimensional subspace, then we would expect 
the remaining directions of spacetime $X^{\alpha}$, (i.e those transverse 
to the flat worldvolume directions $X^{\mu}$ and also to the embedding space), 
to be rotationally invariant. Rotational invariance in $X^{\alpha}$
implies that the metric is diagonal in this subspace and that the
undetermined functions in the metric ansatz depend only on the radial
coordinate $\rho = \sqrt{X_{\alpha} X^{\alpha}}$. 

Isometries, however,
fail to guide us when it comes to dictating the form of the Hermitean
metric $G_{M {\bar N}}$ in the complex space where the holomorphic cycle
is embedded. All we can say is about the Hermitean metric
 is that that it too, along with $H_1$ and $H_2$, must be independent
 of $X^{\mu}$ since we would Lorentz symmetry to be preserved along 
the un-wrapped directions of the M-brane worldvolume. This Lorentz symmetry
also implies that the metric is diagonal in $X^{\mu}$.

A metric incorporating the above symmetries takes the form:
\be
ds^2  = {H_1}^2 \eta_{\mu \nu} dX^{\mu} dX^{\nu} +
2 G_{M {\bar N}} dz^{M} dz^{\bar N} + {H_2}^2
\delta_{\alpha \beta} dX^{\alpha} dX^{\beta}
\label{standard}
\ee
We now describe some of its key features. 

\bit
\item
Comparing this metric to that of a flat M-brane, we see one major
difference. The metric for a wrapped brane too, depends only
on the radial coordinate $r$ in the space transverse to the brane but 
because of the non-trivial worldvolume of this brane
the transverse space is not as simple to define as it was earlier. We 
know that the transverse space now is some combination of what we call 
the relative transverse (or embedding space) directions and the 
overall transverse directions, but a more
exact statement is hard to make, unless we know explicitly the geometry of the 
worldvolume. \\

Depending on how the supersymmetric cycle lies in the 
relative transverse/embedding manifold, it could ofcourse happen that 
$r$ does not depend on all the coordinates in this space. But in order to 
cover all possible cases and for the purposes of making a general ansatz, we 
assume $r$ is a function of all but the overall common
directions.
\item
The Hermitean metric in the above ansatz will have, in general, 
off diagonal components as well. As will be shown later in an explicit 
example, it is these components of the metric which 
allow us to move from a supergravity solution with a smeared intersection 
to one which has a localised intersection. It also allows us to incorporate 
configurations of branes intersecting at angles -- a class of systems not
encompassed by the harmonic function rule
\item
Allowing $r$ to depend on the 
embedding space coordinates has another 
implication also. For supergravity solutions 
constructed via the harmonic function rule, the field strength 
$F_{p + 2} = d A_{p+1}$ 
consisted only of components obtained from the gauge potential by acting $d$ 
in the overall transverse directions, as these were the only ones on which the 
metric was allowed to depend.\\ 

Using the Fayyazuddin-Smith ansatz instead, where the fields are allowed 
to depend on the relative transverse directions as well, we can now apply 
the exterior derivative in these transverse directions also, leading to
previously unknown components for the field strength! 
\item
Moreover, by allowing $r$ to depend on the embedding space coordinates, we 
get rid of the
objection which the harmonic function rule levelled at brane configurations 
with less
than three overall transverse directions. Such configurations too, are
 welcomed into the 
fold of this new ansatz. 
\eit

New lands however, can not be added to the kingdom of supergravity solutions 
without 
paying a price. The battle we must wage now is against non-linear
differential equations.
Earlier, while discussing the flat M-brane solutions, we pointed out
that the functions H were harmonic due to the fact that equations of motion
for the four-form field strength $d*F = 0$ reduced to the flat space Laplace 
equation.
With the more complicated metric we have now, this is no longer the case. The 
undetermined
functions in the metric ansatz are thus harmonic no longer, and obey a more 
complicated 
differential equation, which will in practise be very difficult to solve. 

\section*{New Frontiers}

The fact that a brane configuration preserves supersymmetry implies that 
the supersymmetric variation of the gravitino $\delta_{\chi} \Psi$ 
vanishes in this background if the variation parameter is a Killing spinor. 
If we require this to be true for a given metric, we find a 
set of conditions relating the metric to 
components of the field strength of the supergravity three-form. If  
the resulting metric and four-form also obey  
the contraints $dF=0$ and $d \star F =0$ 
then Einstein's equations are guaranteed to be satisfied and we have determined
the bosonic components of a BPS solution to 11-dimensional supergravity.

Denoting flat (tangent space) indices in 11-dimensional spacetime by
$i,j$ and curved indices by $I,J$, the bosonic part of the action 
for 11d supergravity can be written as
\be
S = \int {\sqrt {- G}} \{
R - \frac{1}{12} F^2 - \frac{1}{432} \epsilon^{I_1....I_{11}}
F_{I_1...I_4} F_{I_5....I_8} A_{I_9..I_{11}}
\}
\ee
and the supersymmetric variation of the gravitino is given by
\be
\delta\Psi_{I} = (\partial_{I}  + \frac{1}{4} \omega_I^{ij} \hat{\Gamma}_{ij}
+ \frac{1}{144}{\Gamma_{I}}^{JKLM}F_{JKLM}
-\frac{1}{18}\Gamma^{JKL}F_{IJKL})\chi. \label{susy}
\ee
Following the logic outlined in the previous section, 
we begin our search for solutions of this
theory by writing down an ansatz for the space-time metric 
following Fayyazuddin and Smith. 

We then enforce supersymmetry preservation 
by setting $\delta \Psi_{I} = 0$. This expression involves 
components of the spin connection which can be calculated
from the metric ansatz using the formula
\be
2 \omega_I^{ij} = e^{iJ}(\partial_I e^j_J - \partial_J e^j_I)
- e^{jJ}(\partial_I e^i_J - \partial_J e^i_I)
- e^{iK}e^{jL}(\partial_K e^l_L - \partial_L e^l_K)e_{lI}
\ee
Since $\chi$ can been expressed as a sum of Fock space states, the
condition $\delta \Psi_{I} = 0$ amounts to a sum of linearly 
independent constraints (one
arising from every Fock state), each of which must be put to
zero seperately. This leads to a set of relations between 
various components of the field strength and the metric. 

The supergravity solution is obtained \cite{wrapM5localised} 
when these conditions are
supplemented by the equations of motion and Bianchi identity for the
four-form field strength $dF=0$ and $d \star F =0$. This is 
the method followed in the preceeding section, to find the bosonic 
solutions of 11-dimensional supergravity for all M-branes wrapping holomorphic
curves such that $F \wedge F = 0$.

\section{Charging forth.}

\subsection*{Membranes}

The Fayyazuddin-Smith ansatz for a metric describing the 
supergravity  background created by an M2-brane wrapping 
a holomorphic curve in $\C^n$ is
\cite{Amherst}:
\be
ds^2 = - H_1^2  dt^{2} + 2 H_1^{-1} g_{M {\bar N}} dZ^{M}
dZ^{\bar N} + H_2^2 \delta_{\alpha \beta} dX^{\alpha} dX^{\beta}.
\label{eq:standard}
\ee
Here $Z^{M}$ are used to denote the $n$ complex coordinates, 
$X^{\alpha}$ span the remaining $(10 - 2n)$ transverse directions 
and a factor of $H_1^{-1}$ is pulled out of the Hermitean metric
for later convenience.

A Hermitean two-form $\omega_g$ associated with the metric is 
defined such that $\omega_g = i {g_{M {\bar N}} dz^{M} \wedge
dz^{\bar N}}$. Supersymmetry preservation imposes a constraint 
on the Hermitean
metric $g_{M {\bar N}}$ and in addition states that\footnote{As 
pointed out in \cite{Amherst}, this 
determines H only up to a rescaling by an arbitrary holomorphic
function} $$H^{-1} = \sqrt{{\rm det} \; g_{M \bar{N}}}$$
where $H \equiv H_1^{-3} = H_2^6$. Further, a set of relations
between components of the metric and field strength is
obtained. These can be solved for the non-zero components of the
four-form. 

Killing spinors of this configuration being Majorana can be
expressed as $$\chi = \alpha + C \alpha^*$$ if C is the charge 
conjugation matrix. Hence, the spinor $\alpha$ is all that is 
needed to completely specify $\chi$

The Bianchi indentiy $dF=0$ is automatically satisfied, but the equation
of motion $d*F = 0$ must be
imposed by hand. This leads to a complicated non-linear differential equation
involving $g_{M {\bar N}}$ and H which must be satisfied by any 
supergravity solution. 

Having thus discussed the structure of the supergravity solutions for
membranes wrapping holomorphic two-cycles in manifolds of various
dimension, we proceed to present the results. 

In some cases, when the holomorphic functions describing the embedding
of the manifold are factorizable, the wrapped brane configuration has 
an intersecting brane interpretation when the curve becomes singular. 
Along with the components of the solution described above, we will also
present in our analysis of each configuration \cite{Amherst}, 
\cite{MeM2}, the intersecting brane 
system  which would arise in the singular limit of this curve, 
if indeed such a singular limit exists.

\newpage

\subsection*{M2 wrapping a 2-cycle in $\C^2$}

\no
\underline{\bf \sf Metric Constraint}\\
\be
\partial (H {\omega}_g) = 0
\label{mc2}\\
\ee

\no
\underline{\bf \sf Four-Form Field Strength}\\

\no
The non-zero components of the four-form field strength are given by:
\bea
F_{0 M \bar{N} \alpha} &=& - \frac{i}{2} \partial_{\alpha} g_{M \bar{N}}, \\
F_{0 M \bar{N} \bar{P}} &=& - \frac{3i}{4} [
\partial_{\bar{P}} g_{M \bar{N}} - \partial_{\bar{N}} g_{M \bar{P}} ]
- \frac{i}{2} [ (\partial_{\bar{P}} {\rm ln} H ) g_{M \bar{N}} - 
(\partial_{\bar{N}} {\rm ln} H ) g_{M \bar{P}} ] \nonumber
\label{fstm2c2}
\eea
and their complex conjugates.\\ 

\no
\underline{\bf \sf Killing Spinors}\\

\no
The eight Killing spinors of this configuration are specified by
\be
\alpha = H^{-1/6} \eta |00>
\ee
such that $\eta$ is a constant spinor obeying the projection condition
\be
\Gamma_{012} \eta = \Gamma_{034} \eta = \eta
\ee

\no
\underline{\bf \sf $d*F = 0$:}\\
\be
\partial^2_{\alpha} (H g_{M \bar{N}} ) +
2 \partial_{M} \partial_{\bar{N}} H = 0
\ee

\no
\underline{\bf \sf Intersecting Brane Limit}\\
\be
\begin{array}[h]{|c|c|cccc|cccccc|}
  \hline
   \; & 0 & 1 & 2 & 3 & 
              4 & 5 & 6 & 7 & 8 & 9 & 10\\
  \hline
  {\bf M2} & \times & \times & \times &  &  &  
               &  &  &  &  &  \\
  {\bf M2} & \times &  &  & \times
               & \times &  &  &  &  &  &  \\
  \hline
\end{array}
\ee\\

\newpage

\subsection*{M2 wrapping a 2-cycle in $\C^3$}

\no
\underline{\bf \sf Metric Constraint}\\
\be
\partial (H {\omega}_g \wedge {\omega}_g) = 0
\label{mc3}
\ee\\

\no
\underline{\bf \sf Four-Form Field Strength}\\

\no
The following expressions, together with their complex conjugates 
give the non-zero components of the field strength.
\bea
F_{0 M \bar{N} \alpha} &=& - \frac{i}{2} 
\partial_{\alpha} g_{M \bar{N}}, \\\nonumber
F_{0 M \bar{N} \bar{P}} &=& - \frac{i}{2} 
[\partial_{\bar{P}} g_{M \bar{N}} - 
\partial_{\bar{N}} g_{M \bar{P}} ]
\label{fstm2c3}
\eea

\no
\underline{\bf \sf Killing Spinors}\\

\no
The four Killing spinors of this configuration 
are specified by 
\be
\alpha = H^{-1/6} \eta |000>
\ee
where $\eta$ is a constant spinor such that
\be
\Gamma_{012} \eta = \Gamma_{034} \eta = \Gamma_{056} \eta = \eta 
\ee

\no
\underline{\bf \sf $d*F = 0$:}\\
\be
\partial_{\alpha}^2 [H \; {\omega}_g \wedge {\omega}_g ] = 0.
\ee

\no
\underline{\bf \sf Intersecting Brane Limit}\\

\be
\begin{array}[h]{|c|c|cccccc|cccc|}
  \hline
   \; & 0 & 1 & 2 & 3 & 
              4 & 5 & 6 & 7 & 8 & 9 & 10\\
  \hline
  {\bf M2} & \times & \times & \times &  &  &  
               &  &  &  &  &  \\
  {\bf M2} & \times &  &  & \times
               & \times &  &  &  &  &  &  \\
{\bf M2}  & \times &  &   &  &  & \times
               & \times &  &  &  &  \\
  \hline
\end{array}
\ee\\
\newpage

\subsection*{M2 wrapping a 2-cycle in $\C^4$}

{\em Since this is a configuration which has only two overall transverse
directions, note that its supergravity solution could not be
obtained using the harmonic function rule.}\\

\no
\underline{\bf \sf Metric Constraint}\\

\no
Supersymmetry requires the metric to obey 
\be
\partial (H \; {\omega}_g \wedge {\omega}_g \wedge {\omega}_g) = 0
\label{mc4}
\ee

\no
\underline{\bf \sf Four-Form Field Strength}\\

\no
The field strength components are the same as the previous case. They are
given by the following expressions and their complex conjugates. 
\bea
F_{0 M \bar{N} \alpha} &=& - \frac{i}{2} 
\partial_{\alpha} g_{M \bar{N}}, \\\nonumber
F_{0 M \bar{N} \bar{P}} &=& - \frac{i}{2} 
[\partial_{\bar{P}} g_{M \bar{N}} - 
\partial_{\bar{N}} g_{M \bar{P}} ]
\label{fstm2c4}
\eea

\no
\underline{\bf \sf Killing Spinors}\\

\no
The two Killing spinors are specified by 
\be
\alpha = H^{-1/6} \eta  |0000 >
\ee
where the constant spinor $\eta$ obeys the projection conditions
\bea
\Gamma_{012} \eta = \Gamma_{034} \eta = \Gamma_{056} \eta =
\Gamma_{078} \eta = \eta
\eea

\no
\underline{\bf \sf $d*F = 0$:}\\

\be
\partial_{\alpha}^2 [H \; {\omega}_g  \wedge {\omega}_g \wedge 
{\omega}_g] + 2 \partial {\bar \partial} 
[H \; {\omega}_g \wedge {\omega}_g] = 0.
\ee

\no
\underline{\bf \sf Intersecting Brane Limit}\\

\be
\begin{array}[h]{|c|c|cccccccc|cc|}
  \hline
   \; & 0 & 1 & 2 & 3 & 
              4 & 5 & 6 & 7 & 8 & 9 & 10\\
  \hline
  {\bf M2} & \times & \times & \times &  &   &  
               &   &   &   &   &   \\
  {\bf M2} & \times &   &   & \times
               & \times &   &   &   &   &   &  \\
{\bf M2}  & \times &   &    &   &   & \times
               & \times &   &   &   &  \\
{\bf M2} & \times &   &   &   &   &   &   & \times
               & \times  &   &  \\
 \hline
\end{array}
\ee\\

\newpage

\subsection*{M2 wrapping a 2-cycle in $\C^5$}

This is the maximal dimensional complex manifold we can have, since it
now spans all ten spatial directions. There is hence no $X^{\alpha}$
and consequently no $F_{0 M \bar{N} \alpha}$ component in the field 
strength. Moreover, there is no $H_2$, so $H$ is defined simply as 
$H = H_1^{-3}$\\

\no
{\em The absence of any overall transverse directions also means that 
the supergravity solution of this
configuration has no counterpart which can be obtained via the
harmonic function rule.}\\

\no
\underline{\bf \sf Metric Constraint}\\
\be
\partial (H \; {\omega}_g \wedge {\omega}_g \wedge {\omega}_g 
\wedge {\omega}_g) = 0
\label{mc5}
\ee

\no
\underline{\bf \sf Four-Form Field Strength}\\

\no
The only non-zero contributions come from:
\be
F_{0 M \bar{N} \bar{P}} = - \frac{i}{2} [
\partial_{\bar{P}} (g_{M \bar{N}}) - \partial_{\bar{N}} (g_{M \bar{P}}) ],
\label{fstm2c5}
\ee
and its complex conjugate, $F_{0 \bar{M} N P}$.\\

\no
\underline{\bf \sf Killing Spinors}\\

\no
The single Killing spinor of this configuration is 
\be
\chi = H^{-1/6} \eta  (|00000> + |11111>)
\ee
where the constant spinor $\eta$ is subject to 
\be
\Gamma_{012} \eta = \Gamma_{034} \eta = \Gamma_{056} \eta =
\Gamma_{078} \eta = \Gamma_{09(10)}\eta = \eta
\ee

\no
\underline{\bf \sf $d*F = 0$:}\\
\bea
g^{M \bar{N}} \partial_{M} H \; 
(\partial_{C} g_{B \bar{N}}  - \partial_{B} g_{C \bar{N}}) 
 = 0\\
g^{M \bar{C}} \partial_{M}  [ H^{- \frac{1}{3}} \; (
\partial_{B} g_{A \bar{C}}  - \partial_{A} g_{B \bar{C}})] = 0
\nonumber \\
g^{N \bar{P}} 3 \partial_{\bar{P}} ( 
\partial_{B} g_{N \bar{C}}  - \partial_{N} g_{B \bar{C}}) = \nonumber \\
g^{N \bar{P}} g^{A \bar{M}} \partial_{\bar{P}} [ g_{A \bar{C}}
\partial_{B} g_{N \bar{M}} + g_{B \bar{M}} \partial_{A} g_{N \bar{C}}
- \partial_{N} (g_{B \bar{M}} g_{A \bar{C}}) ]
\eea

\no
\underline{\bf \sf Intersecting Brane Limit}\\
\be
\begin{array}[h]{|c|c|cccccccccc|}
\hline
   \; & 0 & 1 & 2 & 3 & 
              4 & 5 & 6 & 7 & 8 & 9 & 10\\
\hline
  {\bf M2} & \times & \times & \times &  &  &  
               &  &  &  &  &  \\
  {\bf M2} & \times &  &  & \times
               & \times &  &  &  &  &  &  \\
{\bf M2}  & \times &  &   &  &  & \times
               & \times &  &  &  &  \\
{\bf M2} & \times &  &  &  &  &  &  & \times
               & \times  &  &  \\
{\bf M2} & \times &  &  &  &  &  &  &  &  
	       &\times & \times  \\
\hline
\end{array}
\ee\\

\no \underline{\large \sf \bf Fivebranes}\\

We now turn to supergravity solutions describing fivebranes wrapped on
holomorphic curves. The relevant ansatz for the spacetime metric is:
\be
ds^2 = H^{- 1/3}  \eta_{\mu \nu} dX^{\mu} dX^{\nu}
+ 2 G_{M {\bar N}} dz^{M} dz^{\bar N} +
H^{2/3} \delta_{\alpha \beta} dX^{\alpha} dX^{\beta}.
\ee
where $\mu$ labels the flat coordinates of the fivebrane worldvolume, 
$z^{M}$ are the holomorphic coordinates on the embedding space
$\C^n$ and $\alpha$ takes values in the overall transverse space. 

Supersymmetry preservation has already been used here to fix the 
relative coefficients $H^{- 1/3}$ and $H^{2/3}$. It further dictates 
that, (upto rescaling by an arbitrary holomorphic function), the 
function H is related to the determinant G of the Hermitan metric in a
way that depends on the dimensions of the holomorphic curve and the
complex space into which it is embedded. The Hermitean two-form 
$\omega_G$ associated with the metric in the complex subspace is 
defined such that $\omega_G = i {G_{M {\bar N}} dz^{M} \wedge
dz^{\bar N}}$.

Components of the four-form field strength can be expressed in terms 
of the functions in the metric ansatz by solving a set of constraints
which arise from $\delta \Psi = 0$. Once again, the Killing spinors 
are Majorana and hence of the form
$\chi = \alpha + C \alpha^*$ if C is the charge conjugation matrix. 
The spinor $\alpha$ depends on the particular configuration under study.

The non-linear differential equation
involving $g_{M {\bar N}}$ and H now follows from imposing the 
Bianchi Identity $dF=0$, since the four-form F couples magnetically
to the fivebranes and the roles of the equations of motion and 
Bianchi Identity are consequently interchanged. 

All the supergravity solutions for
fivebranes wrapping holomorphic cycles in manifolds of various
dimension will be of the form outlined above. We present the results
\cite{FS1},\cite{Us},\cite{Amherst},\cite{MeM5} including in addition, the 
intersecting brane system which arises
when the holomorphic cycle becomes singular. Note that such a limit
does not always exist.

\newpage
\subsection*{M5 wrapping a 2-cycle in $\C^2$}

\no
\underline{\bf \sf Metric Constraint}\\
\be
\partial [H^{1/3} \omega_G] = 0.
\label{m5c2}
\ee
where $H^{2/3} = \sqrt{G}$\\

\no
\underline{\bf \sf Four-Form Field Strength}\\

\no
The non vanishing components of the supergravity four-form are:
\bea
F_{M 8 9 (10)} &=& - \frac{i}{2} \partial_{M} H \\
F_{N \bar{M} \beta \gamma} &=& \frac{i}{2} \epsilon_{\alpha \beta \gamma}
\partial_{\alpha} [H^{1/3} G_{N \bar{M}}]
\label{fstm5c2}
\eea\\
and complex conjugates.\\

\no
\underline{\bf \sf Killing Spinors}\\

\no
The eight Killing spinors of this configuration are specified by
\be
\alpha = H^{-1/12} \eta |00>
\ee
where the constant spinor $\eta$ is subject to 
\be
i \Gamma_{0123} \eta = \eta 
\ee

\no
\underline{\bf \sf $dF = 0$:}\\
\be
\partial^2_{\alpha} g_{M \bar{N}}  + 2 \;
\partial_{M} \partial_{\bar{N}} H = 0
\ee

\no
\underline{\bf \sf Intersecting Brane Limit}\\
\be
\begin{array}[h]{|c|cccc|cccc|ccc|}
  \hline
   \; & 0 & 1 & 2 & 3 & 
              4 & 5 & 6 & 7 & 8 & 9 & 10\\
  \hline
  {\bf M5} & \times & \times & \times & \times & \times & \times
               &  &  &  &  &  \\
  {\bf M5} & \times & \times & \times & \times
               &  &  & \times & \times &  &  &  \\
  \hline
\end{array}
\ee

\newpage
\subsection*{M5 wrapping a 2-cycle in $\C^3$}

\no
\underline{\bf \sf Metric Constraint}\\
\be
\partial [H^{- 1/3} \omega_G \wedge \omega_G] = 0.
\label{m5c3}
\ee
and $H = \sqrt{G}$ must hold. \\

\no
\underline{\bf \sf Four-Form Field Strength}\\

\no
The four-form field strength is given by the expressions below, and
their complex conjugates:
\bea
F_{\bar{N} \bar{P} M y} &=& \frac{1}{2}
[\partial_{\bar{P}} (H^{1/3} G_{M \bar{N}}) -
\partial_{\bar{N}} (H^{1/3} G_{M \bar{P}})],\\
F_{M N \bar{P} \bar{Q}} &=& \frac{i}{2} \partial_{y}
[H^{-1/3} (G_{M \bar{Q}} G_{N \bar{P}} - G_{M \bar{P}} G_{N \bar{Q}})]
\label{fstm5c3}
\eea

\no
\underline{\bf \sf Killing Spinors}\\

\no
The four Killing spinors of this configuration are specified by
\be
\alpha = H^{-1/12} \eta |000>
\ee
where the constant spinor $\eta$ obeys the projection condition
\be
i \Gamma_{0123} \eta = \eta 
\ee

\no
\underline{\bf \sf $dF = 0$:}\\
\be
\partial_y^2 (H^{-1/3} \omega_G \wedge \omega_G) \wedge dy 
+ 2 i \bar {\partial} \partial (H^{1/3} \omega_G) = 0
\ee

\no
\underline{\bf \sf Intersecting Brane Limit}\\
\be
\begin{array}[h]{|c|cccc|cccccc|c|}
  \hline
   \; & 0 & 1 & 2 & 3 & 
              4 & 5 & 6 & 7 & 8 & 9 & 10\\
  \hline
  {\bf M5} & \times & \times & \times & \times & \times & \times
               &  &  &  &  &  \\
  {\bf M5} & \times & \times & \times & \times
               &  &  & \times & \times &  &  &  \\
{\bf M5} & \times & \times & \times & \times
               &  &  &  &  & \times & \times &  \\
  \hline
\end{array}
\ee\\

\bigskip
\begin{center}
\noindent\fbox {\noindent\parbox{4.5in}{{\sf \bf Non-Kahler Metrics.}\\

\no
A flat M5-brane can be thought of as being wrapped on a trivial supersymmetric
cycle embedded into a subspace of 11-dimensional spacetime. Take this 
subspace to be $\C^3$, spanned by holomorphic coordinates 
$u, v, w$. Two equations $f = f(u,v,w) = 0$ and $g = g(u,v,w) = 0$ are then 
needed 
to define a holomorphic two-cycle. If these equations are $v = 0$ 
and $w = 0$, the two-cycle in question is simply the complex $u$ plane. 

In the presence of a flat M5-brane with worldvolume $0123u\bar{u}$, the 
spacetime metric is given by: 
$$ds^2 = H^{-1/3} (- dt^2  +  dx_1^2 + dx_2^2 + dx_3^2 + du d{\bar u}) + 
H^{2/3} (dv d{\bar v} + dw d{\bar w} + dy^2) $$
where 
$$H = \frac{{\rm constant}}{(|v|^2 + |w|^2 + y^2)^{3/2}}$$

In general, an M5-brane wrapping a Riemann surface embedded in $\C^3$ is 
expected
to give rise to a metric of the form \cite{Us}:
$$ds^2 = H^{-1/3} (- dt^2  +  dx_1^2 + dx_2^2 + dx_3^2) + 2 G_{M {\bar N}} 
dz^M dz^{\bar N} + H^{2/3} dy^2 $$
Comparing the two expressions and defining a 
re-scaled Hermitean metric $g_{M {\bar N}} = H^{-1/6} G_{M {\bar N}}$, we find
$$2 g_{u {\bar u}} = H^{- 1/2} \; \; , \; \; 2 g_{v {\bar v}} =  H^{1/2}
\; \; , \; \; 2 g_{w {\bar w}} = H^{1/2}.$$ 
It can trivially be seen that this metric is not Kahler; moreover
since a Kahler metric cannot be obtained even by rescaling, 
$g_{M {\bar N}}$ is not warped Kahler either. However, the components 
of this blatantly non-Kahler metric satisfy the following curious relations: 
\be
\partial_u (g_{v {\bar v}} g_{w {\bar w}}) =
\partial_v (g_{u {\bar u}} g_{w {\bar w}}) =
\partial_w (g_{u {\bar u}} g_{v {\bar v}}) = 0.
\ee
In terms of the Hermitean form 
$\omega = i g_{M {\bar N}} dz^M dz^{\bar N}$ associated with the metric, 
this can be re-expressed as follows:
$$\partial [\omega \wedge \omega] = 0 \; \; {\rm but} \; \; \partial \omega 
\neq 0 $$ }}
\end{center}
\bigskip

\subsection*{M5 wrapping a 4-cycle in $\C^3$}

\no
It turns out to be convenient to define a rescaled metric in this case
\be
g_{M {\bar{N}}} \equiv H^{1/3} G_{M {\bar{N}}}
\label{m5c4}
\ee

\no
\underline{\bf \sf Metric Constraint}\\
\be
\partial \omega_g = 0
\ee
and $g = {\rm det} g_{M {\bar{N}}} = H$\\

\no
\underline{\bf \sf Four-Form Field Strength}\\
\bea
F_{8 9 (10) M} &=& \frac{i}{2} \partial_{M} H \\
F_{N \bar{M} \beta \gamma} &=& \frac{i}{2} \epsilon_{\alpha \beta \gamma}
\partial_{\alpha} g_{N \bar{M}}
\label{fieldstrengthfsansatz}
\eea

\no
\underline{\bf \sf Killing Spinors}\\

\no
This configuration has four Killing spinors specified by 
\be
\alpha = H^{-1/12} \eta |000>
\ee
where the constant spinor $\eta$ obeys
\be
\Gamma_{89(10)} \; \eta = -i \eta 
\ee

\no
\underline{\bf \sf $dF = 0$:}\\
\be
\partial^2_{\alpha} g_{M \bar{N}}  + 2 \;
\partial_{M} \partial_{\bar{N}} H = 0
\ee\\

\no
\underline{\bf \sf Intersecting Brane Limit}\\

\no
In the subcases for which an intersecting brane picture exists, this
would take the form:
\be
\begin{array}[h]{|c|cc|cccccc|ccc|}
  \hline
   \; & 0 & 1 & 2 & 3 & 
              4 & 5 & 6 & 7 & 8 & 9 & 10\\
  \hline
  {\bf M5} & \times & \times & \times & \times & \times & \times
               &  &  &  &  &  \\
  {\bf M5} & \times & \times & \times & \times
               &  &  & \times & \times &  &  &  \\
{\bf M5} & \times & \times &  &  
               & \times & \times & \times & \times &  &  &  \\
  \hline
\end{array}
\ee\\

\bigskip
\begin{center}
\noindent\fbox{\noindent\parbox{5in}{{\sf \bf The Harmonic Function
Rule is Safe!}\\

We now show how this new ansatz encompasses and generalises the
harmonic function rule, using the above example to illustrate the
point. The Fayyazuddin-Smith metric ansatz for this system 
can be written as
\be
ds^2 = H^{2/3}  [ H^{-1} \eta_{\mu \nu} dX^{\mu} dX^{\nu}
+ 2 H^{- 1} g_{M {\bar N}} dz^{M} dz^{\bar N} +
\delta_{\alpha \beta} dX^{\alpha} dX^{\beta}] \nonumber
\ee
For ease of comparison, the metric dictated by the harmonic function
rule can also be written using complex coordinates on the embedding
(relative transverse) space. Defining $u = X^2 + i X^3$, $v = X^4 + i
X^5$ and $w = X^6 + i X^7$, this metric takes the form
\bea
ds^2 &=& H_1^{2/3}  H_2^{2/3} H_3^{2/3}
[H_1^{-1}  H_2^{-1} H_3^{-1}(- dX_{0}^2 + dX_{1}^2) 
+ H_1^{-1}  H_2^{-1} du d\bar{u} \nonumber \\ 
&+&  H_1^{-1} H_3^{-1} dv d\bar{v} 
+ H_2^{-1} H_3^{-1} dw d\bar{w}
+ (dX_{8}^2 + dX_{9}^2 +dX_{10}^2)] \nonumber
\eea
Comparing the two expressions, one can immediately see that
\be
H = H_1 H_2 H_3, \;\;\;
2 g_{u {\bar u}} = H_3, \;\;\; 
2 g_{v {\bar v}} = H_2, \;\;\; 
2 g_{w {\bar w}} = H_1 \nonumber
\ee
and all other components of $g_{M {\bar N}}$ are zero. Since
$H_1, H_2$ and $H_3$ are only functions of the overall transverse
coordinates, it follows that H must be too, so $\partial_{M} H =
0$. Substituting in (\ref{fieldstrengthfsansatz}), we find that this
implies $F_{8 9 (10) M} = 0$
and $F_{N \bar{M} \beta \gamma} \propto \eta_{N \bar{M}}$ such that
\bea 
F_{u \bar{u} \beta \gamma} &=& \frac{1}{2} \epsilon_{\beta \gamma \alpha}
\partial_{\alpha} H_3, \; \; \; \; 
F_{u \bar{u} \beta \gamma} = \frac{1}{2} \epsilon_{\beta \gamma \alpha}
\partial_{\alpha} H_2 \nonumber\\
F_{u \bar{u} \beta \gamma} &=& \frac{1}{2} \epsilon_{\beta \gamma \alpha}
\partial_{\alpha} H_1 \nonumber
\eea
in perfect agreement with the results obtained in (\ref{fstm5c3}). 

It should by now be clear, through this explicit discussion, that the
Harmonic Function Rule is the restriction of the Fayyazuddin-Smith
ansatz to cases where H is not allowed to depend on the embedding
space!}}
\end{center}
\bigskip 

\chapter{The Emerald City}
\begin{figure}[ht]
\epsfxsize=6cm
\centerline{\epsfbox{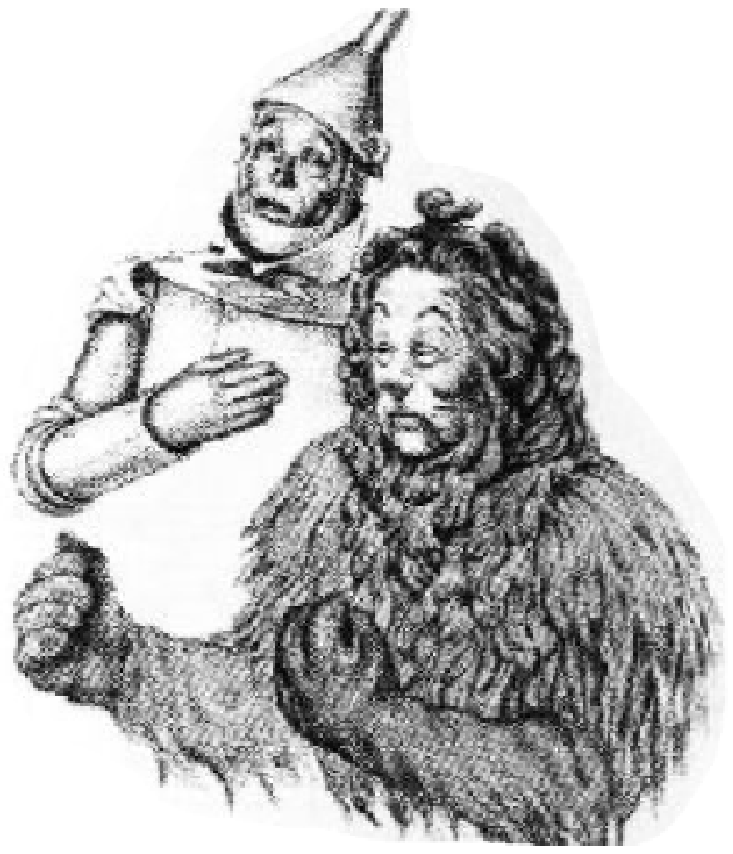}}
\end{figure}

You might have arrived at the doors of the Emerald City, but in order
to get through the gate-keeper and set foot inside the hallowed
portals, you must know the password: {\bf \sf calibrations}. 

Even if this is just another word to you right now, the reason for it
being so revered will become clear when you enter the
shimmering jewelled towers of the Emerald City; in fact once you become
accquainted with calibrations, you will wonder how you ever got along
without them!

\section{The Gate-Keeper}

A calibration is a mathematical construction which enables us to track down
preferred (minimal) submanifolds in a given spacetime. The minimization in
question does not necessarily refer to the volume of the submanifold; 
in fact it is only in purely geometric backgrounds that this is so. As we 
shall soon see, when the background contains a non-vanishing flux, 
calibrations 
pick out minimum energy submanifolds. In 
general, the quantity being minimized tells us about the structure of the 
ambient space-time, so essentially calibrations provide us with a way
of describing a particular background. 

A useful way to discover the calibrations in a particular background
\cite{Cal} is 
to use 'test branes'. The idea is similar to a test particle in 
electrodynamics which is placed in an already existing electric 
field merely to measure its value at a given point. In order to focus
our attention on the properties of the existing background, all
possible distractions are set to zero, i.e  it is assumed that
the test particle is truly just a probe in that it gives 
rise to no fields of its own. 
Similarly, test branes are branes which do not themselves cause 
any deformations of the background in which they are placed, (i.e all
such deformations are neglected) but merely act as probes
for the existing geometry. 

If a test brane is BPS, it takes on a stable shape which must by 
definition, be a supersymmetric cycle in the given background. 
When this background is flat space, we know what these cycles are. 
When the background is a manifold of special holonomy, again we 
know what the cycles are, Berger classified them for us. The problem though 
is to figure out what happens when the background is one created by a charged 
gravitating 
p-brane source! Such a background has a non-trivial flux, and the 
supersymmetric cycles 
in these geometries have not yet been classified. These are the cases we 
will attempt to study 
as we tour the Emerald City.

Requiring the world-volume of a wrapped brane to be supersymmetric determines
or defines the supersymmetric cycles. In the same way, requiring the energy of 
a wrapped brane to be minimal determines the calibrations! 
This is like the age old chicken and egg conundrum $\cdots$ 
who knows what came first? The logic can flow either way. 

\subsection*{Reduced Holonomy}

In the absence of the four-form field strength, a supersymmetric
compactification of M-Theory is only possible on a manifold ${\cal M}$
of reduced holonomy. The reason for this is as follows.
Supersymmetry is only preserved if the supersymmetric variation 
of the gravitino vanishes. 
In a background with no field strength, we can see from (\ref{susy}) 
that setting the gravitino variation to zero is equivalent to
the statement that Killing spinors $\chi$ on the manifold ${\cal M}$
are covariantly constant\footnote{with respect to a torsion-free metric}, 
since \cite{GSW}
\be
\delta_{\chi}\Psi_{I} =
(\partial_{I}  + \frac{1}{4} \omega_I^{ab} \hat{\Gamma}_{ab}) \chi =
D_I \chi = 0,
\ee
where $a = 1,... $ dim ${\cal M}$. From $D_I \chi = 0$, it follows that 
${\cal M}$
is Ricci flat. Furthermore, since 
\be
R_{abIJ} \gamma^{ab} \eta = D_{[J}, D_{I]} \eta = 0
\ee
some generators of the holonomy group $R_{abIJ} \gamma^{ab}$ annihilate 
the vacuum, so the holonomy group of the manifold must be a subgroup of
the maximal SO(dim ${\cal M}$) . 

In other words, for backgrounds where there is no field strength flux, 
supersymmetry
preservation implies the existence of covariantly constant spinors on the
compactification manifold. These exist only if ${\cal M}$ has reduced holonomy 
and is Ricci flat, ensuring that Einstein's field equations are automatically
satisfied by the internal manifold\footnote{A covariantly constant
spinor is always a signal for reduced holonomy, but the converse is
true for Ricci flat manifolds only}. 

For compactifications of string
theory down to four dimensions, these conditions imply that the
manifold ${\cal M}$ should be a Calabi-Yau, ie, a 3 complex dimensional
Ricci flat manifolds with SU(3), rather than SO(6), holonomy; for
4-dimensional compactifications of M-theory, we find instead that
${\cal M}$ is a seven dimensional manifold with $G_2$ holonomy.

\subsection*{Introducing Calibrations}

Calibrations $\phi$ are $p$-forms, which enable us to classify minimal
p-dimensional submanifolds in a particular background space-time \cite{HL} .
The (standard) calibration $\phi$ satisfies 
\be
\int_{{\cal M}_p} \phi \; \leq \; Vol \; ({\cal M}_p) \;\;\;\;\;\;\;
{\rm and} \;\;\;\;\;\;\; d \phi = 0
\ee
for any $p$-dimensional manifold ${\cal M}_p$. In every homology class
there is a manifold $\Sigma_p$ which minimizes the volume, saturating
the above inequality. Since any other manifold ${\Sigma}_{p}^{'}$ in
the same homology class can be written as
\be
{\Sigma}_{p}^{'} = \Sigma_p + \partial {\cal M} _{p + 1}
\ee
Using Stokes' Theorem, we find
\be
\int_{{\Sigma}^{'}} \phi = 
\int_{\Sigma} \phi + \int_{\partial {\cal M}} \phi 
=
\int_{\Sigma} \phi + \int_{{\cal M}} d \phi 
= 
\int_{\Sigma} \phi = Vol \; ({\Sigma}).
\ee
This can be used to prove that
\be
\int_{{\Sigma}^{'}_p} \phi \; = \; Vol \; ({\Sigma}_p) 
\leq \; Vol \; ({\Sigma}_{p}^{'}).
\ee
Hence, integrating the calibration $\phi$ over any $p$-dimensional
submanifold gives the volume of the minimal manifold in that particular
homology class; this minimal manifold is known as a calibrated manifold.

Since we will mostly be dealing with branes of infinite spatial extent,
their volumes will obviously be infinite as well. It thus makes more sense
to express the definition of a calibrating form as follows:
\be
{\cal P}_{{\cal M}_p} (\phi_p) \leq dV_{{\cal M}_p}
\ee
i.e, the pullback of the calibrating form $\phi_p$ onto any $p$-dimensional 
manifold ${\cal M}_p$ is less than or equal to the volume form of this
manifold. 

\subsection*{Berger's Classification.}

For space-times with no background field strength, we have seen that 
the compactification manifold must have special holonomy. There is a 
classification, due to Berger, of the calibrations which can exist on
such manifolds. 

There are several different kinds of calibrations; Kahler calibrations
and special Lagrangian calibrations which exist in any Calabi-Yau space
and exceptional calibrations which exist only in exceptional holonomy spaces.

\bit
\item
In a Calabi-Yau $n$-fold, 
the $2p$-forms $\phi = \frac{1}{p!} \omega^p$ 
which are constructed from the Kahler form $\omega$ are closed and are known
as {\bf Kahler} calibrations.
\item
A Calabi-Yau $n$-fold admits a unique nowhere vanishing ($n,0$)-form, 
$\phi_n$, called the {\bf Special Lagrangian} calibration.
\item
In seven dimensional manifolds, there is a 3-form $\psi$ (and its dual
4-form $*\psi$) which is invariant under the exceptional group $G_2$;
these give rise to three and four dimensional calibrations known as
the {\bf Associative} and {\bf Co-associative} calibrations respectively.
\item
Similarly, in eight dimensional manifolds, there is a
self dual 4-form $\Phi$ which is invariant under $Spin(7)$ and gives
rise to a four dimensional calibration called the {\bf Cayley} calibration.
\eit

\subsection*{Branes and Calibrations}

In the absence of flux, stable branes are those whose world-volumes
are minimized. Given what we have learnt about calibrations, it is obvious that
the volume form on such a brane must be a calibrated form in the ambient
spacetime!

For a membrane wrapped on a holomorphic cycle in $\C^n$,
the volume form decomposes into a trivial one form contribution from the
time direction and the two form associated with the Hermitean metric
in the complex emdedding space of the two-cycle. 
In order for the volume form to be a calibration,
it turns out that the metric on $\C^n$ should in fact be Kahler.

It is hence easy to see why a membrane wrapping a holomorphic two-cycle
in a Calabi-Yau space gives rise to a supersymmetric configuration.

\section{An Audience with the Wizard}

In the presence of a charged brane, the bosonic fields needed to specify
the background are the metric and the flux of the gauge field which
couples to the brane.  
Killing spinors of the brane configuration are determined by the 
metric as well as the field strength and are hence no longer covariantly 
constant. 
Since the coupling of the gauge potential to the brane must now be taken 
into account, it is only natural that the criterion for brane
stability also should change. In fact it turns out that stability
requires now that the energy of the brane be minimized, where the energy is
a measure of not only the volume but also the charge \cite{GCal} . 

{\bf Generalised calibrations} $\phi_p$ are defined such that
\bea
d (A_{p} + \phi_p) &=& 0 \\
{\cal P}_{\Sigma_p} (\phi_p) &\leq& \tilde{dV}_{\Sigma_p}
\label{defgcal}
\eea
So, a generalised calibration $\phi_p$ is a $p$-form whose 
pullback on to 
any $p$-dimensional manifold ${\Sigma_p}$ is less than
or equal to the (curved space) volume form of this manifold. 
Note that $\tilde{dV}$ is used to denote the 
volume form in curved space, as opposed to the flat space volume
form $dV$ used in the definition of a standard calibration.
 
It is clear from the above that a generalised calibration is not closed,
but rather is gauge equivalent to the potential $A_p$ under which the
$(p-1)$brane is charged. In the trivial case where the field strength 
flux vanishes, generalised calibrations reduce to the standard 
calibrations $d\phi = 0$ discussed earlier. 

Recall from our ealier discussion that a calibrated manifold 
i.e, one which saturates the bound, must have minimal
volume. We now proceed to show that an analogous statement holds for 
generalised calibrations as well, except that calibrated manifolds now have 
minimum energy. As we will show later \cite{GCal}, the energy ${\cal E}$ is given by
\be
{\cal E} ({\Sigma}_p) = \int_{\Sigma_p} [ \tilde{dV} + A_p ]
\ee
We start by assuming that $\Sigma_p$ is a calibrated manifold and 
thus saturates the calibration bound 
$$ \int_{\Sigma_p} \phi_p = \int_{\Sigma_p} \tilde{dV} = 
\tilde{Vol}(\Sigma_p ) $$
Since $\Sigma_p$ is minimal, any other generic manifold 
${\Sigma}_{p}^{'}$ in the same homology class can be expressed as 
$\Sigma_p + \partial {\cal M} _{p + 1}$. This implies the following
\be
\int_{{\Sigma}_{p}^{'}} \phi_p = \int_{\Sigma_p}  \phi_p + 
\int_{\partial {\cal M}_{p + 1}} \phi_p 
= \int_{\Sigma_p} \phi_p  + 
\int_{{\cal M}_{p + 1}} d \phi_p
\ee
From the definition of generalised calibrations, we have
$d A_{p} = -d \phi_p$, 
\bea
\int_{{\Sigma}_{p}^{'}} \phi_p &=& \int_{\Sigma_p} \phi_p - 
\int_{{\cal M}_{p + 1}} d A_p 
= \int_{\Sigma_p} \phi_p - 
\int_{\partial {\cal M}_{p + 1}} A_p \nonumber \\
&=& \int_{\Sigma_p} \phi_p +  \int_{\Sigma_p} A_p - 
\int_{{\Sigma}_{p}^{'}} A_p
\eea
Hence, 
\be
\int_{{\Sigma}_{p}^{'}} [\phi_p + A_p] = \int_{\Sigma_p} [\phi_p + A_p] 
\label{aaa}
\ee\\

\no
For the manifold ${\Sigma}_{p}^{'}$, 
$$\int_{{\Sigma}_{p}^{'}} \phi_p \leq \int_{{\Sigma}_{p}^{'}}  \tilde{dV} 
\;\;\;\;\; \Rightarrow \;\;\;\;\;
\int_{{\Sigma}_{p}^{'}} [\phi_p + A_p] \leq {\cal E}({\Sigma}_{p}^{'})$$
However, for the calibrated manifold the inequality is saturated, and 
$$\int_{\Sigma_p} [\phi_p + A_p] = {\cal E}(\Sigma_p)$$
\no
Hence (\ref{aaa}) reduces simply to the statement 
$$ {\cal E} ({\Sigma}_p) \leq {\cal E} ({\Sigma}_{p}^{'}) $$
which says that the calibrated manifold is the one which minimizes its energy
in a given holonomy class. 

Having seen the way things are in the Emerald City, let us now try and 
understand, as far as we can, the reasons {\bf why} they are so. 

\subsection*{Charged Branes}

The Lagrangian density for a $p$-brane charged under a potential ${\cal A}
_{p+1}$ is
\be
{\cal L} = - \sqrt{- {\rm det} \; h_{ab}} - \cal{P}(A)
\ee
where $h_{ab}$ is the induced metric on the world-volume, and 
$${\cal P}({\cal A}) = \epsilon^{a_0 a_1 \dots a_p} \partial_{a_0} X^{{\mu}_0} 
\partial_{a_1} X^{{\mu}_1} \dots 
\partial_{a_p} X^{{\mu}_p} {\cal A}_{{\mu}_0 {\mu}_1 \dots {\mu}_p}$$ 
is the pullback of the spacetime (p+1)-form gauge potential onto the brane. 
In what follows, we will consider only those spacetimes for which the 
component 
$G_{0I}$ of the metric vanishes, if I is a spatial index. 
Seperating the purely spatial part of the induced metric by defining
\be
g_{ij} \equiv \partial_i X^{\mu} \partial_j X^{\nu} G_{\mu \nu} 
\ee 
we can express ${\rm det} \; h_{ab}$ as:
\be
{\rm det} \; h_{ab} = (\partial_t X^{\mu} \partial_t X^{\nu} G_{\mu
\nu}) {\rm det} \; g_{ij} 
\ee
Choosing now the gauge $X^0 = t$, the Lagrangian density takes the form
\bea
{\cal L} = &-& \sqrt{(G_{tt} + \partial_t X^{I} \partial_t X^{J}
G_{IJ}) {\rm det} \; g_{ij}}
\nonumber \\
&-& \epsilon^{i_1 \dots i_p} \partial_{i_1} X^{{I}_1} \dots 
\partial_{i_p} X^{I_p} {\cal A}_{t I_1 \dots I_p}
\nonumber \\
&-& \epsilon^{i_1 \dots i_p} \partial_t X^M \partial_{i_1} X^{{I}_1} \dots 
\partial_{i_p} X^{I_p} {\cal A}_{M I_1 \dots I_p}
\eea
Since the the canonical momentum $p_I$ is given by the expression
\bea
p_K &=& \frac{\partial {\cal L}}{\partial (\partial_t X^K)} \nonumber \\
&=& - \frac{\partial_t X^J G_{JK} 
-  \epsilon^{i_1 \dots i_p} \partial_{i_1} X^{{I}_1} \dots 
\partial_{i_p} X^{I_p} {\cal A}_{K I_1 \dots I_p}}{\sqrt{- {\rm det} 
\; h_{ab}}}
\eea
it follows that the Hamiltonian density ${\cal H} = p_I \partial_t X^I - 
{\cal L}$ is
\be
{\cal H} = \sqrt{- {\rm det} \; h_{ab}} - \frac{\partial_t X^J G_{JK} 
+ \epsilon^{i_1 \dots i_p} \partial_{i_1} X^{{I}_1} \dots 
\partial_{i_p} X^{I_p} {\cal A}_{t I_1 \dots I_p}}{\sqrt{- {\rm det}
\; h_{ab}}} 
\ee
Restricting ourselves to static configurations, we can set 
$\partial_t X^J = 0$ to get 
\be
{\cal H} = \sqrt{- {\rm det} \; h_{ab}} 
+ \epsilon^{i_1 \dots i_p} \partial_{i_1} X^{{I}_1} \dots 
\partial_{i_p} X^{I_p} {\cal A}_{t I_1 \dots I_p}
\ee
For notational convinience we now introduce the $p$-form $A_p$ such that
\be
A_{I_1 \dots I_p} \equiv {\cal A}_{t I_1 \dots I_p}
\ee
This allows us to define $\Phi$, the worldvolume pull-back 
of the space-time gauge potential, as follows
\be
\Phi \equiv * {\cal P}({\cal A}) = 
\epsilon^{i_1 \dots i_p} \partial_{i_1} X^{{I}_1} \dots 
\partial_{i_p} X^{I_p} {\cal A}_{I_1 \dots I_p}
\ee
Finally, we are in a position to write down the energy ${\cal E}$, 
associated with the $p$-brane as follows
\bea
{\cal E} &=& 
\int d^p\sigma [ \sqrt{- G_{tt}} \sqrt{{\rm det} \; g_{ij}} +
\Phi] \nonumber \\ 
&=& \int d^p\sigma [ \tilde{dV} + A ]
\eea

\subsection*{Spatial Isometries}

Typically, only those directions along the brane worldvolume which are wrapped 
on a supersymmetric curve have a non-trivial space-time embedding; the 
remaining directions are flat\footnote{In the notation adopted here, 
these are the $X^{\mu}$}. Had its entire world-volume been flat, the 
brane would be a 1/2 BPS object; wrapped branes however, generically 
break more than half the supersymmetry. Since the preserved
supercharges depend on the geometry of the supersymmetric cycle, it is
the wrapped directions of the brane world-volume which play a
essential role in this analysis whereas flat directions contribute trivially. 

We can choose static gauge along the flat directions of the brane's
world-volume. Assume there are $l$ such directions, we then set 
$X^i = \sigma^i$ for $i = 1 ,\dots,l$ and find that as a result, 
the determinant of the induced metric on the
worldspace factorises such that
\be
{\rm det} \; h_{ab} = - G_{tt} \; {\rm det} \; G|_{l \times l} \; {\rm
det} \; g_{rs} \equiv \nu^2_l \; {\rm det} \; g_{rs} 
\ee
where ${\rm det} \; G|_{l \times l}$ denotes the restriction of the
determinant of the bulk space metric to the $l$ spatial directions
which are isometries of the system, and ${\rm det} \; g_{rs}$ is the
determinant of the induced metric on the Mp-brane in directions 
$\sigma^{l+1} \dots \sigma^p$. Also, the pullback of the gauge
potential ${\cal P}({\cal A})$ can be expressed in static gauge as
\bea
{\cal P}({\cal A}) &=& 
\epsilon^{i_1 \dots i_p} \partial_{i_1} X^{{I}_1} \dots 
\partial_{i_p} X^{I_p} A_{I_1 \dots I_p} \nonumber \\
&=& 
\epsilon^{i_{l+1} \dots i_p} \partial_{i_{l+1}} X^{{I}_{l+1}} \dots 
\partial_{i_p} X^{I_p} A_{1 2 \dots l I_{l+1} \dots I_p}
\eea
Hence the energy of this configuration is given by
\be
{\cal E} = \int d^p\sigma[ \nu_l \; \sqrt{{\rm det} \; g_{rs}} 
+ 
\epsilon^{i_{l+1} \dots i_p} \partial_{i_{l+1}} X^{{I}_{l+1}} \dots 
\partial_{i_p} X^{I_p} A_{1 2 \dots l I_{l+1} \dots I_p}]
\ee
Due to the infinite extent of the brane, this energy too will be infinite. 
It thus makes 
sense to define the energy per unit $l$-volume
\be
{\cal E}_l = \int d^{p-l}\sigma [ \nu_l \; \sqrt{{\rm det} \; g_{rs}} + 
\epsilon^{i_{l+1} \dots i_p} \partial_{i_1} X^{{I}_{l+1}} \dots 
\partial_{i_p} X^{I_p} A_{1 2 \dots l I_{l+1} \dots I_p}].
\ee

\subsection*{Supersymmetry Preservation}

In order to be a Killing spinor, $\xi$ must satisfy the projection condition
\cite{GCal}
\be
\frac{1}{\sqrt{{\rm det} h}} \epsilon^{a_0 a_1 \dots a_p} \partial_{a_0} 
X^{{\mu}_0} 
\partial_{a_1} X^{{\mu}_1} \dots 
\partial_{a_p} X^{{\mu}_p} \Gamma_{{\mu}_0 {\mu}_1 \dots {\mu}_p} \xi
= \xi 
\ee
Assuming that the configuration is static, this condition takes the
form
\be
\frac{1}{\gtt \sqrt{{\rm det} h}} \epsilon^{a_1 \dots a_p} 
\partial_{a_1} X^{{\mu}_1} \dots 
\partial_{a_p} X^{{\mu}_p} {\Gamma_0} \Gamma_{{\mu}_1 \dots {\mu}_p} \xi
= \xi 
\ee
Defining 
\be
\gamma
\equiv 
\epsilon^{a_1 \dots a_p} 
\partial_{a_1} X^{{\mu}_1} \dots 
\partial_{a_p} X^{{\mu}_p} \Gamma_{{\mu}_1 \dots {\mu}_p} 
\ee
we can re-write the above condition as
\be
[1 - \frac{1}{\gtt \sqrt{{\rm det} h}} {\Gamma_0} \gamma] \xi
\equiv [1 - \hat{\Gamma}]\xi = 0
\label{gammahat}
\ee
where $(\hat{\Gamma})^2 = 1$ since $\gamma^2 = 1$ and 
$\Gamma_0^2 = - G_{tt}$. 

\subsection*{Establishing a bound}

The obvious statement that the square of a quantity is positive
definite can be used to establish a general bound on $\phi|_{\xi}$. 

Consider$[1 - \hat{\Gamma}]\xi$, acting on a generic configuration.
Even if the system is not supersymmetric, the following will
obviously hold
\be
\xi^{\dagger} [1 - \hat{\Gamma}]^{\dagger} [1 - \hat{\Gamma}] \xi
\geq 0 \nonumber
\ee
Since $\hat{\Gamma}$ is Hermitean and the spinor $\xi$ has 
been normalised such that $\xi^{\dagger} \xi = \gtt \sqrt{{\rm det} h}$, 
we find 
\be
\gtt \sqrt{{\rm det} h} [1 - \xi^{\dagger} \hat{\Gamma} \xi]
\geq 0 \nonumber
\ee
Inserting the expression for $\hat{\Gamma}$ from (\ref{gammahat}), 
we can re-write the above as
\be
\gtt \sqrt{{\rm det} h} -  \xi^{\dagger} {\Gamma_0} \gamma \xi \geq
0 \nonumber 
\label{xxx}
\ee
Since $\gtt \sqrt{{\rm det} h} = {\rm Vol}_{\xi}$ is the (world-space
dual of the) curved
space volume form, this implies that in terms of the $p$-form
\be
\phi \equiv * \; \xi^{\dagger} {\Gamma_0} \gamma \xi 
\label{const}
\ee
which is the world-space dual of $\xi^{\dagger} {\Gamma_0} \gamma \xi $,
the inequality (\ref{xxx}) reduces to 
\be
{\rm Vol} (\cal M) \geq {\cal P}_{\cal M} (\phi)
\ee
But this is merely the definition of a 
generalised calibration, (\ref{defgcal})! So (\ref{const}) 
is actually a way to explicitly construct
generalised calibrations from Killing spinors! 

\subsection*{Supersymmetry Algebra}

The supersymmetry algebra is motivated in \cite{GCal} from considerations
of $\kappa$-symmetry. It turns out that, 
\bea
\bar{\xi} \{\bar{Q}, Q \} \xi &=&  
\int d^{p}\sigma [ \gtt \sqrt{{\rm det} h} -  \bar{\xi} {\Gamma_0}  
\gamma \xi] \nonumber \\
&=& \int d^{p}\sigma [ {\cal H} - \Phi - * \phi] \nonumber \\
&=& H - \int d^{p}\sigma [A + \phi]
\eea
So, $\int d^{p} \sigma [A - \phi]$ constitues a central extension to the 
superalgebra; as such, it must be a topological term. This implies that
\be
d (A + \phi) = 0
\ee
which was one of the defining statements about generalised calibrations.

\section{The Great and Powerful Oz has Spoken!!}

The Great Oz now convinces us of his greatness by showing us a simple
 and elegant way
to reproduce the results we worked so hard for, in our trek down the Yellow
Brick Road. Had we never walked that road, he argues, we could still
have reached the same conclusions, if we only had the knowledge he has. All
the supergravity solutions which we found after a long and arduous
 trek, the Wizard now reproduces in the comfort of his Emerald City
 quarters. The trick about to be performed will consist of a few short steps,
carried out in quick succession, which result in the construction 
of a supergravity solution for any M-brane wrapped on a holomorphic
curve and will replicate the expressions obtained from the
Fayyazuddin-Smith analysis employed by us during our Yellow
Brick Road travels.\\

{\sf ``A supergravity solution''}, says the Wizard, {\sf `` consists of
a metric, which may be expressed in terms of undetermined functions, 
as long as we also
present the equations which these functions must satisfy and in addition
the four-form field strength of the supergravity three-form''}. He will
now prove to us his unquestionable superority, he says, through a point by 
point comparison of our earlier slower approach to his fast and
elegant analysis.

Recall that we started out with the Fayyazuddin-Smith ansatz for a
metric, and then obtained relations between the undetermined functions 
$H_1, H_2$ and det $G_{M {\bar N}}$ in the ansatz by appealing to the 
constraints which arose from setting $\delta \Psi = 0.$ The Wizard replicates 
these relations as follows:\\

\no
Given the Fayyazuddin-Smith ansatz for the metric in a particular
supergravity background, the functions $H_1$ and $H_2$ can be
eliminated in favour of a single function $H$.
\bit
\item
For a membrane, $H_1 = H^{-1/3}$ and  $H_2 = H^{1/6}$, 
\item
And for a fivebrane  $H_1 = H^{-1/6}$ and  $H_2 = H^{1/3}$. 
\eit
\no
Note in both cases, the analogy with the flat brane solutions.\\

\no
The relation between the determinant of the Hermitean metric and H can be 
obtained using the following simple relations which also follow 
from comparison with the flat brane case: 
\bit
\item
For a membrane on a holomorphic cycle, the determinant of the full 
11-dimensional metric is always $H^{2/3}$. The determinant G
of the Hermitean metric in a manifold of complex dimension $n$ must
be given by $G = H^{(2n -6)/3}$ in order for this to hold\footnote{Bear 
in mind that all such relations hold only upto multiplication by an arbitrary 
holomorphic function}.
\item
For a fivebrane wrapped on a holomorphic two-cycle, the 
determinant of the full 11-dimensional metric is $H^{4/3}$. The
determinant G of the $n$ dimensional Hermitean metric must therefore 
be $G = H^{(2n -6)/3}$.
\item
For a fivebrane wrapped on a holomorphic four-cycle, the 
determinant of the full 11-dimensional metric
is still $H^{4/3}$. The
determinant G of the $n$ dimensional Hermitean metric however is now 
$G = H^{(2n - 12)/3}$.
\eit

{\sf ``So that''}, says the Wizard, {\sf ``takes care of that. True,''} 
he adds, {\sf ``the non-linear differential equation involving the 
Hermitean metric and 
H will have to be written out, but this follows simply from 
requiring $d*F = 0$ for membranes, and $dF=0$ for fivebranes. So, once
I present you with the expressions for components of F, my work is
done.''} 

Once again, he reminds us, where we had to wade through many different
constraints, one for each Fock state of the Killing spinor, and solve
them simultaneously to arrive at expressions for F, the Wizard will 
reproduce these in the wink of an eye. {\sf ``Pay attention,''} he says, {\sf 
``because this is where the magic really happens''.} The Wizard now claims 
that components of the field strength can be calculated simply by acting 
the exterior derivative on the volume form of the wrapped M-brane in question!
Noticing the awe-struck expression on our faces, the gleam in his eye
deepens and he elaborates as follows:

{\sf ``It is the BPS bound,''} says he, {\sf ``which steps in to make 
matters so
 simple. Since this bound is saturated by a supersymmetric brane, 
the mass of such a brane must be equal to its charge. Equally, one could
say that the pull-back of the space-time gauge potential on 
to a supersymmetric brane is equal to the volume form of the brane, 
by virtue of the BPS condition. As you doubtless remarked during your 
tour of the Emerald City, the
generalised calibration corresponding to a particular stable
brane is (gauge) equivalent to the 
spacetime gauge potential under which the brane in charged.  

``Hence, the generalised calibration for a 
a stable (and hence BPS) brane is given simply by its volume form! 
Components of the field strength can thus be determined by using the fact that
the calibrated form corresponding to a wrapped brane is equivalent to
the gauge potential to which the brane couples electrically''}.\\

For our edification, and to convince us beyond a shadow of a doubt, 
this procedure will now be illustrated in detail for each of the 
M-brane configurations we came across earlier, at a time when we were 
ignorant of calibrations.\\ 

\subsection*{M2 wrapping a 2-cycle in $\C^n$}

Start with writing down the Fayyazuddin-Smith metric 
in the background of M2-brane wrapping 
a holomorphic curve in $\C^n$:
\be
ds^2 = - H^{-2/3} dt^{2} + 2 H^{1/3} g_{M {\bar N}} dz^{M}
dz^{\bar N} + H^{1/3} \delta_{\alpha \beta} dx^{\alpha} dx^{\beta}.
\label{eq:standard}
\ee
The $n$ holomorphic coordinates are denoted by $z^{M}$ and
${\alpha} = 2n + 1 , \dots 10.$ 
A factor of $H_1^{-1}$ has been pulled out of the Hermitean metric 
to facilitate comparison with the expressions found earlier. 

We know from the BPS condition that the calibrating form $\Phi$ of the 
M2-brane must be identical to its volume form and can hence 
be read off directly from the metric
\bea
\Phi &=& i H^{-1/3} g_{M {\bar N}}
dt \wedge dz^M \wedge dz^{\bar{N}}\\
&=& dV_{0} \wedge \phi_{M \bar{N} }
\eea
Since $F_4$ is the electric field strength for the M2-brane, it can be 
calculated using $F_4 = d \Phi$ to yield the expressions
in (\ref{fstm2c2}) through (\ref{fstm2c2}).

\subsection*{M5 wrapping a 2-cycle in $\C^2$}

When an M5-brane wraps a holomorphic 2-cycle in $\C^2$,
the relevant ansatz for the spacetime metric is:
\be
ds^2 = H^{- 1/3}  \eta_{\mu \nu} dX^{\mu} dX^{\nu}
+ 2 G_{M {\bar N}} dz^{M} dz^{\bar N} +
H^{2/3} \delta_{\alpha \beta} dX^{\alpha} dX^{\beta}
\label{m5on2cycle}
\ee
where $z^{M}$ are coordinates on $\C^2$, $\alpha$ takes values $8,9,10$
and $\mu$ runs over $0,1,2,3$. 
The harmonic function H is
related to the determinant G of the Hermitan metric by ${\sqrt G} = 
H^{2/3}.$
Since the calibrating form $\Phi$ of the BPS M5-brane is identical to its
volume form, we can read it off directly from the metric to obtain 
\bea
\Phi &=& i H^{-2/3} G_{M {\bar N}}
dt \wedge dX^1 \wedge dX^2 \wedge dX^3 \wedge
dz^M \wedge dz^{\bar{N}}\\
&=& dV_{0123} \wedge \phi_{M \bar{N} }
\label{vfm22}
\eea
We can now calculate $F_4 = * d F_7 = * d \Phi$ and find the same expressions
as in (\ref{fstm5c2}).

\subsection*{M5 wrapping a 2-cycle in $\C^3$}

When the M5-brane is wrapped on a holomorphic curve embedded in $\C^3$,
the metric takes the form:
\be
ds^2 = H^{- 1/3}  \eta_{\mu \nu} dX^{\mu} dX^{\nu}
+ G_{M {\bar N}} dz^{M} dz^{\bar N} +
H^{2/3} dy^2.
\label{m5on2cycle3}
\ee
where $z^{M}$ now span $\C^3$, $y$ is the single overall transverse direction.
and the harmonic function H is related to the determinant of the Hermitean
metric by $H = \sqrt{G}$.

In this background, the wrapped M5-brane is calibrated by the 
volume form
\bea
\Phi &=& H^{-2/3} G_{M {\bar N}} dt \wedge dX^1 \wedge dX^2 \wedge dX^3 \wedge
dz^M \wedge dz^{\bar{N}}\\
&=& dV_{0123} \wedge \phi_{M \bar{N} }
\label{vfm23}
\eea
The field strength can be calculated using 
$F_4 = * d F_7 = * d \Phi$. This yields the same expressions
as in (\ref{fstm5c3}).

\subsection*{M5 wrapping a 4-cycle in $\C^3$}

For an M5-brane wrapped on a 4-cycle $\Sigma_4$
in $\C^3$, the metric takes the form:
\be
ds^2 = H_1^{2}  \eta_{\mu \nu} dX^{\mu} dX^{\nu}
+ 2 G_{M {\bar N}} dz^{M} dz^{\bar N} +
H_2^{2} \delta_{\alpha \beta} dX^{\alpha} dX^{\beta}
\label{m5on4cycle}
\ee
where $\mu=0,1$ labels the unwrapped directions, $z^{M}$ are
holomorphic coordinates in $\C^3$ and $\alpha$ takes values 8,9, and 10.
The determinant of the Hermitean metric is given by $G = H^{-4/3}$\\

The generalised calibration $\Phi$ for this wrapped M-brane is the 
same as its volume form, hence the only non-vanishing component of
of $\Phi$ is the following:
\bea
{\Phi}_{01 M N {\bar P} {\bar Q} } &=& H^{-1/3}
(G_{M {\bar P}} G_{N {\bar Q}} - G_{M {\bar Q}} G_{N {\bar P}}) \\
&=& dV_{01} \times {\phi}_{M N {\bar P} {\bar Q}} \nonumber
\eea
Using the fact that $F_4 = * d \Phi$ we can once again calculate 
the components of the four-form, obtaining the same results as in
(\ref{fstm5c34}). 

\section{Lifting the curtain.}

In this awe-inspiring show where supergravity solutions of wrapped
M-branes make their appearance with such grace and speed, it is easy 
to get side-tracked and ignore the one major sleight of hand; all the 
information obtained via the Fayyazuddin-Smith method has been duplicated 
here $\cdots$ with the conspicuous omission of the metric constraint!

It is with this slight over-sight that the Great Oz betrays himself as 
a mere magician rather than a real Wizard. His failure to constrain 
the metric in any way, leaves us with a very rich structure, but no 
clue where to apply it.
Having now pointed out the Wizard's short-coming, which doubtless he
was hoping we would not catch on to, we are almost as far away 
from an answer as we were before. 

Without any further knowledge of the Hermitean metrics 
we can allow, it seems that a journey home, or for that matter
anywhere else, is currently
out of the question. As a last resort, we turn to the truly magical 
Ruby Slippers to show us the way. 

\chapter{The Ruby Slippers}

\begin{figure}[ht]
\epsfxsize=7cm
\centerline{\epsfbox{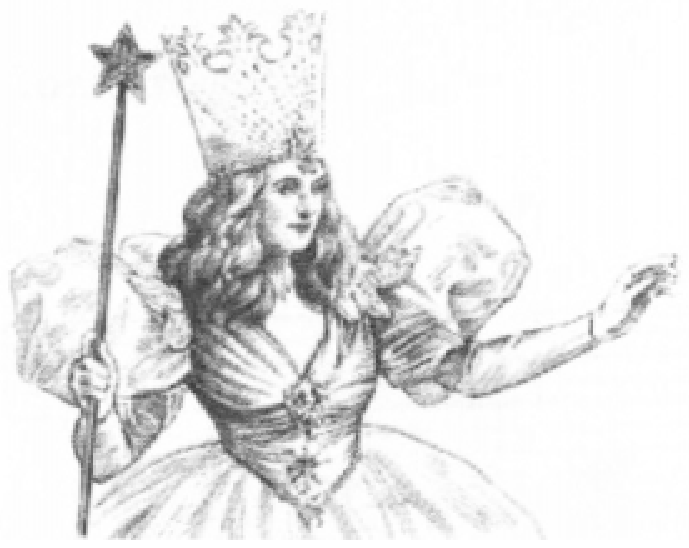}}
\end{figure}

We have seen, in our journey down the Yellow Brick Road, how to 
construct the supergravity solution for a wrapped M-brane by looking 
for bosonic backgrounds which admit Killing spinors. Having set 
the gravitino to zero, we have made sure that the supersymmetric variations 
of the bosonic
fields vanish identically and it is left only to impose that the supersymmetry
variation of the gravitino vanish as well. We require this to be true
for our metric ansatz, when the variation parameter in the supersymmetry
transformation is a Killing spinor. If in addition, the Bianchi identity and 
equations
of motion for the field strength are also satisfied, the metric and four-form
we have obtained are guaranteed to satisfy Einstein's equations, furnishing
a bosonic solution to 11-dimensional supergravity.

Once we arrived in the Emerald City, we learnt that there was an alternate 
way in 
which this problem could be solved. Since the generalised calibration 
corresponding 
to a wrapped M-brane is gauge equivalent to the gauge potential to which the 
brane 
couples electrically, the field strength $F= dA = d\phi$, then immediately 
follows 
once we are given a suitable calibration. For membranes, this field strength 
is the four-form
we are used to seeing in our supergravity solutions; for five-branes, the 
field strength is 
a seven-form and we have to dualise it in order to obtain the familiar 
four-form. 
Supersymmetry requirements fix the undetermined functions in the metric 
ansatz in 
terms of the Hermitean metric $G_{M \bar{N}}$ 
which obeys a non-linear differential equation that follows from $d*F=0$.

This procedure is by far simpler than the previous one... however there
is a catch! Since generalised calibrations which can exist in a background
with non-zero flux have not yet been classified, there is no comprehensive
(or even partial) list from which we can pick a suitable calibration $\phi$, 
from
which to construct a supergravity solution. 

\section*{The Constraint}

A persistent feature of the wrapped brane supergravity solutions we saw
along the Yellow Brick Road is a constraint on the metric in the subspace 
where the supersymmetric cycle is embedded. This constraint in turn restricts 
the 
$(p+1)$-form potential to which the brane couples. Since the 
potential is gauge equivalent to the generalised calibration, we find that 
in fact the metric constraint can alternately be viewed as a condition which
determines generalised calibrations in the given background. 

This condition can be expressed as a constraint on the Hodge dual, with 
respect to the embedding space, of the generalised calibration in that space. 
When written this way, it shows clearly that calibrated 
(supersymmetric) cycles in a particular submanifold of space-time, are 
not specified completely by the submanifold and in fact have a non-trivial 
dependence on the surrounding spacetime as well. This is to
be expected because unlike the case for F=0 when supersymmetric cycles
depended only on geometry, we do expect now that the field strength flux
(and through it, the remaining non-complex directions of spacetime) will
make their presence felt and play their part in determining the calibrated
cycles which can exist in a given subspace. 

Expressing the results in this form also enables us to unify the M2 and M5
brane analysis and to show that the constraint can be expressed perfectly 
generally 
and arises for all M-branes embedded holomorphically into a subspace of 11 
dimensional spacetime, such that $F \wedge F = 0$.

\newpage

\no
\underline{\sf \bf The Magic of the Ruby Slippers}\\

{\sf \bf 
\no
In 11-dimensional backgrounds with non-zero four-form flux, 
a class of generalised calibrations in the embedding space {\cal M} is given 
by the 
$2m$-forms $\phi_{2m}$, for
\bea
\begin{array}{ccc}
\phi_{2m} & = & (\underbrace{\omega \wedge \omega \wedge ..... \omega \wedge 
\omega})\\
 &  & m
\end{array}
\eea
if the following constraint holds:
\be
\partial *_{\cal M} [ \phi_{2m} |G^{'}|^{1/2m} ] = 0.
\label{fcond}
\ee
Here, $G^{'}$ denotes the determinant of the metric restricted to directions 
transverse 
to the embedding space, and the Hodge dual is taken within the embedding 
space.}

\bigskip
\begin{center}
\noindent\fbox {\noindent\parbox{4.5in}{{\sf \bf The Hodge Dual of the 
Hermitean Form.}\\

If $\omega$ is the two-form associated with the Hermitean metric on a manifold 
${\cal M}$ of complex dimension $n$, and $*$ denotes the Hodge dual on this 
space, 
then we have the following:
\bea
{* \omega}_{{\bar P}_2 \dots {\bar P}_n Q_2 \dots Q_n}
&=& \sqrt{{\rm det } g} \; {\epsilon}^{M_1}_{\; \; {\bar P}_2 \dots {\bar P}_n}
{\epsilon}^{\bar{N}_1 \; \; }_{Q_2 \dots Q_n} \omega_{M_1 \bar{N}_1} 
\nonumber \\
&=& \sqrt{{\rm det } g} \; g^{M_1 {\bar S_1}} g^{R_1 \bar{N_1}} 
\tilde{\epsilon}_{\bar{S_1} {\bar P}_2 \dots {\bar P}_n}
\tilde{\epsilon}_{R_1 Q_2 \dots Q_n} i  g_{M_1 \bar{N}_1} \nonumber \\
&=& i \sqrt{{\rm det } g} \; g^{R_1 {\bar S_1} } 
\tilde{\epsilon}_{\bar{S_1} {\bar P}_2 \dots {\bar P}_n}
\tilde{\epsilon}_{R_1 Q_2 \dots Q_n}
\eea
Substituting now the expression for the inverse metric
\be
g^{R_1 {\bar S_1}} = \frac{1}{(n-1)!}  \frac{1}{\sqrt{{\rm det}g}} 
\tilde{\epsilon}^{\bar{S_1} {\bar S}_2 \dots {\bar S}_n}
\tilde{\epsilon}^{R_1 R_2 \dots R_n} g_{R_2 {\bar S}_2} \dots 
g_{R_n {\bar S}_n}
\ee
we find that
\bea
\begin{array}{ccc}
i^{(n-2)} * \omega & = & (\underbrace{\omega \wedge \omega 
\wedge ..... \omega \wedge 
\omega})\\
 &  & (n - 1)
\end{array}
\eea

As a 'check' of the above, note that wedging both sides with $\omega$ gives 
the 
identity 
\bea
\begin{array}{ccccc}
\omega \wedge * \omega & = & 
(\underbrace{\omega \wedge \omega \wedge ..... \omega \wedge 
\omega}) & = & {\rm Vol} ({\cal M}) \\
 &  & n & &
\end{array}
\eea}}
\end{center}
\bigskip

\section{Click the Heels ...}

\subsection*{M2-branes}

Supergravity solutions for a class of BPS states corresponding to 
wrapped membranes were dicussed in \cite{MeM2}. In keeping with the logic 
that holomorphicity implies supersymmetry, the M2-branes were wrapped on 
holomorphic cycles in complex subspaces of varying dimension $n$.

We will work with the following (standard) ansatz for the spacetime metric:
\be
ds^2 = - H^{-2/3}  dt^{2} + 2 G_{M {\bar N}} dz^{M}
dz^{\bar N} + H^{1/3} \delta_{\alpha \beta} dX^{\alpha} dX^{\beta},
\label{eq:standard}
\ee
where $z^{M}$ are $n$ holomorphic coordinates and 
$X^{\alpha}$ span the $(10 - 2n)$ transverse directions. 

Inorder to express the constraint on the generalised calibrations
in each case, we resort to (\ref{fcond}). Since the M2-branes wrap
only two-cycles, the calibrating form in the complex space 
is simply the Hermitean form associated with the metric, 
$\omega = i G_{M {\bar N}} dz^{M} \wedge dz^{\bar N}$. The restricted 
determinant $|G^{'}|$ is also simple to calculate and we find that
\be
|G^{'}| = H^{-2/3}  (H^{1/3})^{10 - 2n} =  (H^{1/3})^{8 - 2n}
\ee

The constraints now follow immediately. For a membrane wrapping a
holomorphic curve in $\C^n$, (\ref{fcond}) dictates that
\be
\partial *_{\C^n} [H^{(4 - n)/3} \omega_G] = 0
\ee
This can be explicitly written out as follows:
\bea
\partial [H^{2/3} \omega_G] &=& 0 \;\;\; {\rm for} \; n=2 \nonumber\\
\partial [H^{1/3} \omega_G \wedge \omega_G] &=& 0 \;\;\; {\rm for} \; 
n=3\nonumber\\
\partial [\omega_G \wedge \omega_G \wedge \omega_G] &=& 0 \;\;\; {\rm for} \; 
n=4 \nonumber\\
\partial [H^{- 1/3} \omega_G \wedge \omega_G \wedge \omega_G \wedge \omega_G] &=& 0 
\;\;\; 
{\rm for} \; n=5
\label{m2metrics}
\eea
Note that these constraints reproduce (\ref{mc2}) - (\ref{mc5})

\subsection*{M5-branes}

The five-brane configurations do not fit this easily into a pattern, and 
will just have to be considered one by one. 

We start with the two M5-branes which wrap a holomorphic 
two-cycle in $\C^n$. For both these systems, the calibrating form
is the Hermitean two-form associated with the metric, as long as it
is subject to the proper constraint. In order to work out the constraint,
we need to first calculate $\sqrt{|G^{'}|}$, where $G^{'}$ is the determinant
of the metric in the directions transverse to $\C^n$.\\  

\no
\underline{\bf \sf M5 wrapping a 2-cycle in $\C^2$}

\no
From the metric (\ref{m5on2cycle}) it is clear that 
\be
|G^{'}| = (H^{-1/3})^4 (H^{2/3})^3 = H^{2/3}
\ee
so, the constraint (\ref{fcond}) for this system is
$$
\partial *_{\C^2} [H^{1/3} \omega_G] = \partial [H^{1/3} \omega_G] = 0 
$$
which agrees with (\ref{m5c2}).\\

\no
\underline{\bf \sf M5 wrapping a 2-cycle in $\C^3$}

\no
The metric (\ref{m5on2cycle3}) allows us to read off 
\be
|G^{'}| = (H^{-1/3})^4 H^{2/3} = H^{- 2/3}
\ee
so the constraint in this case takes the form
$$
\partial *_{\C^3} [H^{-1/3} \omega_G] = 
\partial [H^{-1/3} \omega_G \wedge \omega_G] = 0 
$$
which reproduces (\ref{m5c3}).\\

We now turn to the last configuration on our list, and 
the only one which involves a brane wrapping a four-cycle. The calibrating
form is now the square of the Hermitean two-form, $\omega_G \wedge \omega_G$
and in order to impose the relevant constraint, we need to compute 
$|G^{'}|^{1/4}$.\\ 

\no
\underline{\bf \sf M5 wrapping a 4-cycle in $\C^3$}

\no
In this case, we can see from the metric (\ref{m5on4cycle}) that
\be
G^{'} = (H^{-1/3})^2 (H^{2/3})^3 = H^{4/3}
\ee
and the constraint on the calibration is thus
$$
\partial *_{\C^3} [H^{1/3} \omega_G \wedge \omega_G] = 
\partial [H^{1/3} \omega_G] = 0.
$$
in perfect agreement with (\ref{m5c4}).

Until the explicit forms of these constraints were found 
\cite{FS1}, \cite{Us}, \cite{MeM2}, \cite{MeM5}, the Fayyazuddin-Smith
construction was assumed to apply only to Kahler metrics. 
While (warped) Kahler metrics \cite{Amherst} 
definitely solve all the above constraints they by no means 
exhaust the available options, as we will show in the following section.
So, by assuming that the metric on the 
embedding space is Kahler, we are 
in fact restricting ourselves unnecesarily and losing out on a wealth of 
possibilities. 

\section{Count to Three....}

\no
\underline {\bf \sf Satisfying $\partial (\omega \wedge \omega) = 0$}\\

Since $\omega_{M \bar{P}}  = i G_{M \bar{P}}$, the above constraint 
can be written in component form as 
\be
\partial_{[R} (G_{M \bar{P}} G_{N] \bar{Q}}) = 0
\ee
Upon contraction with the inverse metric $G^{N \bar{Q}}$ this gives
\bea
2 (n - 3) [\partial_R G_{M \bar{P}} - \partial_M G_{R \bar{P}}]
+G_{M \bar{P}} \; \partial_R ln G  
= \nonumber\\ 
2 G^{N \bar{Q}}[G_{M \bar{P}} \; \partial_N G_{R \bar{Q}} -
G_{R \bar{P}} \; \partial_N G_{M \bar{Q}}] + G_{R \bar{P}} \; \partial_M ln G
\label{dg2=0}
\eea
Here, $n$ denotes the complex dimension of the manifold which has
$G_{M \bar{N}}$ as its Hermitean metric. Contracting again 
with $G^{M \bar{P}}$ we find the relation\footnote{
There is an overall factor of (n-2) multiplying this equation, but that can 
be cancelled once we note that on manifolds of complex dimension $n \leq 2$, 
the five-form $\partial (\omega \wedge \omega)$ would vanish identically! 
Similar, 
obviously non-zero factors also occur in the analysis of the remaining 
constraints.}
\be
\partial_R ln G = 2  G^{N \bar{Q}} \partial_N G_{R \bar{Q}}
\ee
which can be substituted into (\ref{dg2=0}) to give
\be
(n - 3) [\partial_R G_{M \bar{P}} - \partial_M G_{R \bar{P}}] = 0
\ee
This equation can be satisfied in two ways: 
\bit
\item
Either $n = 3$, in which case $\partial (\omega \wedge \omega)$ = 0 is a 
non-trivial 
requirement,
\item
Or we must have a Kahler metric, so that $\partial (\omega \wedge \omega)$ 
vanishes
as a result of $\partial \omega = 0.$\\
\eit

\no
\underline {\bf \sf Satisfying $\partial (\omega \wedge \omega \wedge 
\omega) = 0$}\\

Employing the same procedure as in the previous case, we write the constraint
out in component form
\be
\partial_{[U} (G_{M \bar{Q}} G_{N \bar{R}} G_{P] \bar{S}} ) = 0 
\ee
and contract it with $G^{P \bar{S}} G^{N \bar{R}}$ to obtain
\bea
(n - 4) [\partial_U G_{M \bar{Q}} - \partial_M G_{U \bar{Q}}] \; 
+  G_{M \bar{Q}} \partial_U ln G \; = \nonumber\\ 
2 G^{N \bar{R}}[G_{M \bar{Q}} \partial_N G_{U \bar{R}} - 
G_{U \bar{Q}} \partial_N G_{M \bar{R}}]
+ G_{U \bar{Q}} \partial_M ln G 
\label{dg3=0}
\eea
Contracting once more with the inverse metric $G^{M \bar{Q}}$ 
we arrive at the relation
\be
\partial_U ln G = 2  G^{M \bar{Q}} \partial_M G_{U \bar{Q}}
\ee
which can be substituted into (\ref{dg3=0}) to give
\be
(n - 4) [\partial_U G_{M \bar{Q}} - \partial_M G_{U \bar{Q}}] = 0
\ee
Hence, it is only in a four-complex dimensional manifold that 
the constraint 
$\partial (\omega \wedge \omega \wedge \omega) = 0$ can be satisfied
{\it without} having $\partial \omega = 0$. 

In particular this implies that 
$\partial (\omega \wedge \omega \wedge \omega) = 0$
cannot be satisfied by a Hermitean two-form (of a non-Kahler metric) which 
obeys 
$\partial (\omega \wedge \omega) = 0$. This is in agreement with the 
analysis of 
the previous constraint, where it was found that 
$\partial (\omega \wedge \omega) = 0$ is a non-trivial constraint only in 
three complex dimensions.\\ 

\no
\underline {\bf \sf Satisfying $\partial (\omega \wedge \omega \wedge 
\omega \wedge \omega ) = 0$}\\

Writing the constraint out in component form,
\be
\partial_{[A} (G_{B \bar{C}} G_{D \bar{E}} G_{F \bar{H}} G_{I] \bar{J}}) = 0 
\ee
and contracting with $G^{B \bar{C}} G^{D \bar{E}} G^{F \bar{H}}$, leads to the
expression
\bea
2 (n - 5) [\partial_A G_{I \bar{J}} \partial_I G_{A \bar{J}}] 
+ 3 G_{I \bar{J}} \partial_A ln G = \nonumber \\
6 G^{B \bar{C}}
[G_{I \bar{J}} \partial_B G_{A \bar{C}} - G_{A \bar{J}} \partial_B 
G_{I \bar{C}}] + 3 G_{A \bar{J}} \partial_I ln G 
\label{dg4=0}
\eea
This, when contracted further with $G^{I \bar{J}}$, yields the relation
\be
2 G^{B \bar{C}} \partial_B G_{A \bar{C}} = \partial_A ln G
\ee
Substituting the above into (\ref{dg4=0}), we find that
\be
(n - 5) [\partial_A G_{I \bar{J}} - \partial_I G_{A \bar{J}}] = 0
\ee
This statement implies that in order to satisfy 
$$\partial (\omega \wedge \omega \wedge \omega \wedge \omega) 
= 0$$ 
non-trivially, we must have a manifold with complex dimension $n = 5$; in 
all other dimensions, this constraint can only be satisfied by a Kahler 
metric.

\section{... and You're Home!}

So, to summarize, what we learnt in the Emerald City was how to
construct a supergravity solution for a wrapped M-brane by looking
simply at the isometries of the configuration and using the technology
of generalised calibrations. This analysis leads to an almost complete
picture -- the one missing piece in the puzzle is a constraint on
generalised calibration. This constraint as we have now seen, can be  
generated by applying a remarkably simple rule (\ref{fcond}) to the 
spacetime in question. We have also seen that it is solved only by 
two possible classes of calibrations; those corresponding to metrics 
which are either Kahler, $\partial \omega = 0$, or co-Kahler, 
$\partial * \omega = 0$.

Now that you can construct the supergravity solution
for an M-brane wrapped on any holomorphic cycle with a click of your
heels, perhaps it is time to venture futher into lands as yet
unknown. Countless wrapped M-brane configurations exist which have not
been considered here. The most obvious extension would be to look at
M-branes on holomorphic cycles such that $F \wedge F \neq 0$; an
example is provided by an M5 wrapping a holomorphic four-cycle
embedded in a four-complex dimensional manifold. Branching out
further, one could consider M-branes wrapping Special Lagrangian
cycles. It would be interesting to see if an analogous constraint
arises on the embedding space metric in those cases and to explore its
implications for the corresponding calibrations. 

Whether or not the Ruby Slippers will be of any use to us in these new
journeys remains to be seen. Certainly they have served us well in our
little adventure here, and have provided us with plenty of material
about which we can now ``{\em sit, and think some more}''.   
\begin{figure}[ht]
\epsfxsize=6cm
\centerline{\epsfbox{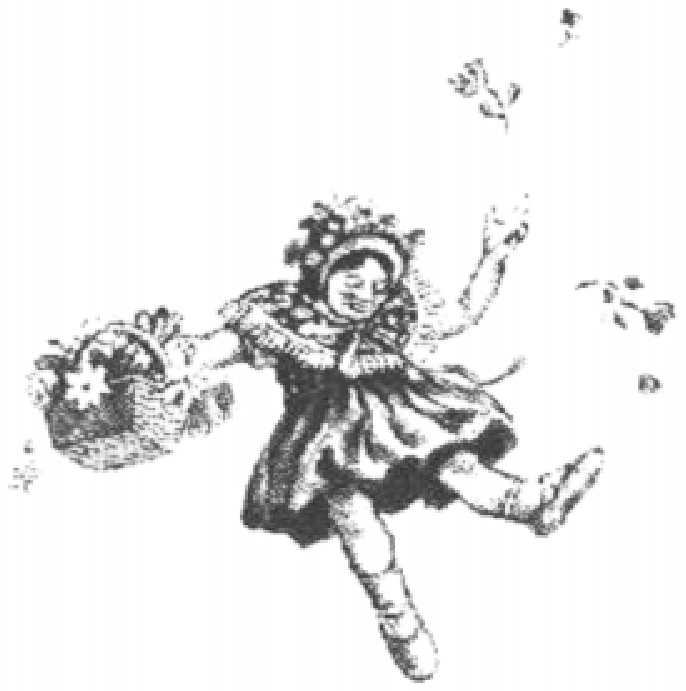}}
\end{figure}

\end{document}